%% file: main.tex
\documentclass[%
 reprint,
 prl,
superscriptaddress,
 amsmath,amssymb,
 aps,
]{revtex4-1}

\usepackage{comment}
\usepackage{dsfont}
\usepackage{graphicx}
\usepackage{dcolumn}
\usepackage{bm}
\usepackage{color}
\usepackage{epsfig}
\usepackage[colorlinks=true, linkcolor=blue, citecolor=blue, urlcolor=blue]{hyperref}
\usepackage{natbib}
\usepackage{float}
\usepackage{footmisc}
\usepackage{slashed}
\usepackage[dvipsnames]{xcolor}
\usepackage{aas_macros}
\definecolor{lightred}{rgb}{1,0.5,0.5}
\definecolor{lightgreen}{rgb}{0.5,1,0.5}
\definecolor{lightblue}{rgb}{0.5,0.5,1}
\definecolor{lightcyan}{rgb}{0.5,0.75,0.75}
\definecolor{lightmagenta}{rgb}{0.75,0.5,0.75}
\definecolor{customgreen}{rgb}{0.494,1,0.502}

\usepackage{soul}

\newcommand{\meV}{\mathinner{\mathrm{meV}}}
\newcommand{\eV}{\mathinner{\mathrm{eV}}}
\newcommand{\keV}{\mathinner{\mathrm{keV}}}
\newcommand{\MeV}{\mathinner{\mathrm{MeV}}}

\newcommand{\be}{\begin{equation}}
\newcommand{\ee}{\end{equation}}
\newcommand{\ba}{\begin{aligned}}
\newcommand{\ea}{\end{aligned}}

\begin{document}

\title{Gradient-Produced Neutrinos}

\author{Erwin H.~Tanin}
\email{ehtanin@stanford.edu}
\affiliation{Leinweber Institute for Theoretical Physics at Stanford, Department of Physics, Stanford University, Stanford, California 94305, USA}
\author{Yikun Wang}
\email{yikunw@jhu.edu}
\affiliation{The William H.~Miller III Department of Physics and Astronomy, The Johns Hopkins University, Baltimore, Maryland, 21218, USA}
\begin{abstract}
   Sufficiently strong electric fields can produce charged‑particle pairs via the Schwinger effect. We argue that steep matter‑density gradients, as can arise in neutron star interiors, would analogously produce neutrino–antineutrino pairs. We then discuss observational signatures of these gradient‑produced (anti-)neutrinos and how they could provide new probes of neutron‑star structure and baryon-dense QCD.
\end{abstract}
\maketitle

Strong background fields can destabilize the vacuum and spontaneously produce particle-antiparticle pairs. The best-known example is electron-positron production in a static and uniform electric field, known as the Schwinger effect \cite{1931ZPhy...69..742S,Schwinger:1951nm,1969JETP...30..660N,Schneider:2014mla}. Variants of the Schwinger effect have been studied in a wide range of setups involving particles other than the electron in non-uniform and/or time-dependent electromagnetic fields, as well as other (non-)Abelian gauge fields  \cite{Gelis:2015kya, Aleksandrov:2016lxd,Gavrilov:1996pz,Ruf:2008ahs,Fedotov:2022ely}. While the critical fields for an unsuppressed rate is beyond current laboratory reach, analogous phenomena may occur in astrophysical settings where extreme conditions arise naturally.

Heuristically, vacuum pair-production occurs when a background field can separate virtual particle-antiparticle pairs by distances greater than their Compton wavelengths. While gauge fields are usually considered for this effect, the heuristic condition can arise in more general backgrounds. Here we focus on inhomogeneous vector backgrounds $j_{\rm ext}^\mu$ \footnote{A scalar $\phi$ interacting with a fermion $\psi$ via Yukawa interaction of the form $\phi\bar{\psi}\psi$, for example, does not meet the conditions, since a background scalar gradient $\bm{\nabla}\phi$ exerts the same force on $\psi$ and its antiparticle and therefore does not separate them.} coupled to fermionic (axial)vector-currents since they provide a direct generalization of Schwinger effect beyond gauge fields \footnote{We expect, however, that the underlying mechanism is broader and can be extended to other classes of backgrounds and couplings. A systematic exploration of vacuum pair-production and its connection to beyond the Standard Model scenarios is postponed for future work \cite{FutureWork}}. Within the Standard Model (SM), neutrinos propagating in matter acquire an effective vector potential from coherent forward scatterings. In baryon-asymmetric matter, this potential has opposite sign for a neutrino and its antineutrino, suggesting that a sufficiently steep matter-density gradient can drive neutrino-antineutrino pair production, in analogy with the Schwinger effect. 

The state of cold, baryon-dense matter in the interior of a neutron star (NS) remains an open question \cite{Fukushima:2010bq,Rajagopal:1999cp}. Many viable models feature sharp density variations arising from first-order phase transitions \cite{Baym:2017whm,Alvarez-Castillo:2017qki}, chemical stratification due to accretion or nuclear burning \cite{Chamel:2008ca}, or local charge conservations \cite{Glendenning:1992vb,Hempel:2009vp,Belvedere:2012uc}. We point out that such sharp density/composition jumps imply steep matter-potential gradients and can lead to significant neutrino-antineutrino production \footnote{Neutrino pair-production by matter-potential gradients in NS was previously studied in Refs.~\cite{1990PhRvL..64..115L,kiers1997coherent,Kachelriess:1997cr,koers2005perturbative}. Those works considered typical NS density gradients with characteristic length scales $n_n/|\nabla n_n|\sim 10\text{ km}$ and concluded that gradient-produced neutrinos have no discernible effects on an NS.  Unlike these earlier works, we consider NS interiors with a sharp density/composition jump, in which case the neutrino pair-production rate can be many orders of magnitude larger.}. We compute the pair-production rate and show that these ``gradient-produced" neutrinos can have \textit{observable} effects on NS cooling. Conversely, their imprint on NS cooling curves could provide a new observational probe of NS interior and of QCD at high baryon densities, a challenging regime for first-principle lattice calculations \cite{Nagata:2021ugx}.

In the remainder of this \textit{Letter}, we develop these ideas in a simplified framework.  Our aim in this initial study is to provide a clear proof of principle to motivate detailed and realistic studies of NS phenomena that incorporate (anti)neutrinos sourced by sharp density gradients. We use natural units with $c=\hbar=k_B=1$.

\paragraph*{\textbf{Generalized Schwinger Effect.---}}  Consider a fermion $\psi$ of mass $m$ coupled to an external current $j_{\rm ext}^\mu$,
\begin{align} \label{eq:Lag}
   - \mathcal{L}_{\rm int}=j_{\rm ext}^\mu\bar{\psi}\gamma_\mu\left(c_V-c_A\gamma_5\right)\psi,
\end{align}
where $c_V$ and $c_A$ are dimensionless couplings. The vector $j_{\rm ext}^\mu$ need not be a gauge or fundamental field. It could be the gradient of a scalar/pseudoscalar $\partial^\mu\phi$, the divergence of a tensor $\partial_\nu S^{\mu\nu}$, or other contractions that yield a vector. For simplicity, we take the external current to be static and purely timelike, $\left<j^\mu_{\rm ext}\right>=(V(\bm{x}),0,0,0)$, however, with a non-trivial spatial dependence. 

Consider helicity states with $h = \pm 1$, the fermion dispersion relation in the WKB limit ($V$ varying slowly in a de~Broglie wavelength) reads $E_{\pm}^h =  c_V V \pm \sqrt{ (|\bm{p}| - c_A h V )^2   + m^2}$, where $E$ is the frequency (instead of energy). Pair production becomes possible when the spatial variation of $V$ allows a positive-frequency (particle) branch and a negative-frequency (antiparticle) branch to cross. The original Schwinger setup of an electron in a constant and uniform electric field $-\partial_zA^0$ corresponds to a purely vector coupling, $c_V=1$ and $c_A=0$, and $V=-eA^0$. A sufficiently strongly spatially varying $A^0$ allows the frequency of a positron at a location to match that of an electron at another location, enabling pair production. The pair-production rate is exponentially suppressed if these points are separated more than a Compton wavelength $m_e^{-1}$.

The dispersion relation simplifies in the massless limit, where the left- and right-handed fermions, $\psi_{L,R}=(1\mp\gamma_5)\psi/2$, decouple, \begin{align} \label{eq:dism}
    E_{L,R}^h &= \mp\, h |\bm{p}| +(c_V\pm c_A)V\quad {\rm for}\quad m \to 0.
\end{align}
They are related to the four solutions in the massive case, labeled by the $\pm$ index, as $E_{L,R}^h= E_{\mp\,{\rm sgn}( h|\bm{p}| -  c_A V)}^h |_{m \to 0}$. Within each chiral sector, a particle and an antiparticle of opposite helicities can be pair-produced. Further details are given in the Supplemental Material Sec.~\ref{sec:apprate}.

\paragraph*{\textbf{The Pair-Production Rate.---}} 
The spatial inhomogeneity of the background field may vary from \textit{gradual} to \textit{sharp}. To capture both limits and the smooth transition between them, we assume a one-dimensional tanh profile for the background potential (Sauter potential) 
\begin{align} \label{eq:V}
    V(\bm{x})=  \frac{1}{2}\Delta V\tanh \left(\frac{z}{l}\right).
\end{align}
Here, $\Delta V$ and $l$ set the strength and length scale of the potential step, respectively. We take $\Delta V>0$ without loss of generality. A necessary condition for pair production is that the potential be \textit{supercritical}, namely that the shift between the two asymptotic regions $\Delta V=V(z\rightarrow +\infty)-V(z\rightarrow-\infty)$ is sufficiently large to connect positive- and negative-frequency solutions.

A supercritical potential $V$ can mix the positive- and negative-frequency modes of the Dirac equation. Upon canonical quantization, this translates to the creation and annihilation operators at $z\rightarrow-\infty$ and $z\rightarrow+\infty$ regions being different. The number of produced particle-antiparticle pairs at a given phase-space point, $N({E,\bm{p}_{\perp}})$, where $\bm{p}_\perp=\bm{p}-p_z\hat{z}$, can be expressed in terms of the Bogoliubov coefficients relating these operators. These coefficients can be computed by solving a scattering problem in the potential $V(z)$ \cite{nikishov1970pair,Gitman:1977ne,Soffel:1982pm,Ambjorn:1982bp,Tanji:2008ku}, yielding $N({E,\bm{p}_{\perp}}) = 1- r ({E,\bm{p}_{\perp}})^{-1}$, where $r ({E,\bm{p}_{\perp}})$ is the reflection probability of a particle incident from $z\rightarrow-\infty$ toward the potential step. Consider the phase space confined within a time interval $T$ and a transverse area $A = \int dx dy$, the particle production rate per transverse area reads 
\begin{align}\label{eq:genrateperarea}
  \dot{n}_{\perp} \equiv \frac{N_{\rm tot}}{T A}
  \approx &\sum_{i=L,R} \int \frac{dE d^2\bm{p}_{\perp}}{(2\pi)^3} \frac{\sinh ( \pi \, l\, p_{i,z}^+)}{ \sinh \frac{\pi\, l}{2}(\Delta V_{i} +  p_{i,z}^+ -  p_{i,z}^-  ) }\nonumber\\
  &\times \frac{\sinh ( \pi \, l \, p_{i,z}^- )}{\sinh \frac{\pi\, l}{2}( \Delta V_i - p_{i,z}^+ +  p_{i,z}^-  )},
\end{align}
where $p_{i,z}^{\pm} \equiv \sqrt{ ( E \mp \Delta V_{i}/2 )^2 - p_{\perp}^2 }$, and $\Delta V_{i=L,R} = |c_{V} \pm c_{A}| \Delta V$ is the potential depth for each chiral sector. Because our application focuses on ultrarelativistic neutrinos, we have adopted the small-mass approximation. In this regime, the two chiral sectors decouple, and the supercriticality condition for production is trivially satisfied. Further details and derivation of the results are given in~Supplementary~Sec.~\ref{sec:apprate}, \ref{sec:solve}, and \ref{sec:pairproductionderivation}.

The pair-production rate per area, $\dot{n}_\perp$, simplifies in certain limits. The largest $\dot{n}_\perp$ at a fixed $\Delta V$ occurs for a sharp interface, $l \ll \Delta V^{-1}$, where 
\begin{align}\label{eq:rateperareasharp}
 \dot{n}_{\perp} \approx 
  \frac{\log 4-1 }{60\, \pi^2}  \Delta V^3 \sum_{h=\pm 1} |c_V-h\, c_A|^3\ 
  ({\rm sharp}).
 \end{align}
Here, the transition width $l$ drops out, and the rate scales as $\Delta V^3$, consistent with earlier results for $c_V =1, c_A = 0$~\cite{Chervyakov:2009bq,Chervyakov:2011nr}.

In the opposite, gradual limit~$l \gg \Delta V^{-1}$, the rate per area (in $l\lesssim \Delta V/m^2$ and zero-mixing limits) becomes 
\begin{align}\label{eq:rateperareagradual}
& \dot{n}_{\perp}  \approx
\frac{\Delta V^2 l^{-1}}{16\pi^3}
\sum_{h=\pm} (c_V-h\,c_A)^2\ ({\rm gradual}).
\end{align}
This gradual limit is related to the Schwinger limit where the potential gradient $g \equiv  \Delta V/2l $ is held fixed as $\Delta V, l \to \infty$.  In that case, production occurs over a large spatial volume, and the production rate is often expressed in terms of the rate per unit volume. The above expression reproduces the prefactor, $\dot{n}_\perp/2 l\approx  g^2/8\pi^3$ \footnote{With a tanh/Sauter potential, the pair production occurs mainly within the spatial region $-l\lesssim z \lesssim l$, where the potential gradient is sizable and approximately uniform. For $|z|\gtrsim l$, the tanh gradient asymptotes to zero. Thus the production rate per volume can be estimated from the production rate per area as $\dot{n} \approx \dot{n}_{\perp}/(2 l)$.}, dependence of the pair-production rate per unit volume in the Schwinger limit with $c_V=1,c_A=0$, $\dot{n}  \approx
\frac{g^2}{8 \pi^3}  e^{-\pi m^2 g^{-1}}$ \cite{sauter1931behavior,heisenberg1936folgerungen,schwinger1951gauge}\footnote{We note that our pair-production rates differ from those of, \textit{e.g.}, Refs.~\cite{Kusenko:2001gb,Koers:2004pj,dvornikov2014creation}, whose leading-order rates are proportional to the particle's mass squared. The difference lies in the backgrounds assumed: static \& inhomogeneous in our case and dynamical \& homogeneous in theirs.}. The Schwinger pair-production rate is exponentially suppressed for $l \gtrsim \Delta V/m^2$. More generally, an axial coupling $c_A\neq 0$ introduces a chiral mixing, which may become important as we increase $l$, possibly before the exponential suppression kicks in. Meanwhile, the chiral mixing can be safely neglected in the sharp limit \footnote{In~Supplementary~Material~Sec.~\ref{sec:apprate}, \ref{sec:solve}, and \ref{sec:pairproductionderivation}, we provide derivations of the production rate neglecting the chiral mixing while keeping the mass dependence in each chiral sector. The result reproduces the exponential suppression behavior in the gradual limit, although the chiral mixing would introduce important corrections as it becomes relevant for sufficiently large $l$. For $l$ small enough to be in the sharp limit, however, the chiral mixing can be safely neglected, as discussed in~Supplementary~Material~Sec.~\ref{sec:solve}. Further dedicated study is needed to capture the full mass dependence of the rate for $c_A\ne 0$.}.

\begin{figure}[t!]
    \centering
\includegraphics[width=1.0\linewidth]{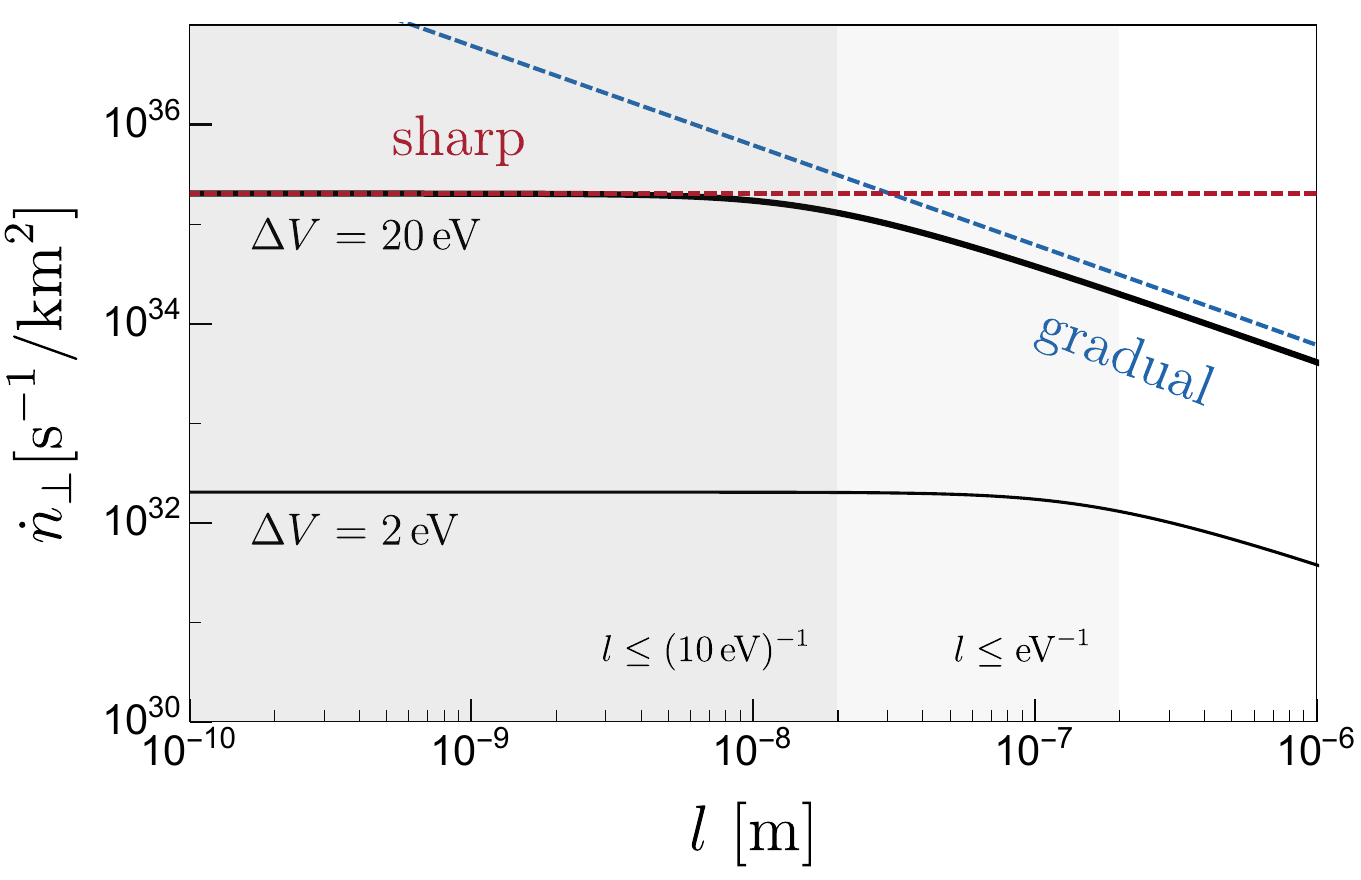}
    \caption{Pair-production rate per unit area $\dot{n}_{\perp}$ by a chiral  ($c_V = c_A = 1/2$) Sauter potential, Eqs.~\eqref{eq:Lag} and \eqref{eq:V}. The thicker (thinner) black solid curve is the numerically evaluated $\dot{n}_{\perp}$ for a Sauter potential with $\Delta V=20\eV$ ($\Delta V=2\eV$). The red (blue) dashed curve shows the approximate rate for $\Delta V = 20\,$eV in the sharp (gradual) limit, which is valid for $l \lesssim (\gtrsim) (\Delta V/2)^{-1}$, as shaded in light gray for the sharp limit.}
    \label{fig:lumi}
\end{figure}

Neutrino interactions with NS matter are described at low energies by the neutral-current four-fermion operator of the form in~Eq.~\eqref{eq:Lag}, with $\psi\rightarrow \psi_\nu$, $c_V=c_A=1/2$ for Dirac neutrinos and $c_V=0,c_A=1$ for Majorana neutrinos. In what follows, we focus on the Dirac case;  the Majorana case is discussed in the~End~Matter. Coherent forward scattering in non-relativistic NS matter with mass density $\rho$ generates an effective potential for neutrinos. For a neutron-rich matter, this potential is $V=30\eV[\rho/(10^{15}\text{ g/cm}^3)]$  \cite{Wolfenstein:1977ue,Notzold:1987ik}.  In general, $V$ is flavor- and NS-phase-dependent, but typically lies in the same ballpark. This potential acts oppositely on left-handed neutrinos and their antineutrinos (attractive for the former and repulsive for the latter), so a sufficiently steep gradient can spontaneously produce neutrino-antineutrino pairs. In~Fig.~\ref{fig:lumi}, we show the rate per area, $\dot{n}_{\perp}$, as a function of $l$ for $c_V = c_A = 1/2$. The solid black curves are obtained for a couple of $\Delta V$ via numerical integration of Eq.~\eqref{eq:genrateperarea}. The colored dashed curves show the sharp and gradual approximations, which match the numerical results well for $l\lesssim (\Delta V/2)^{-1}$ and $l\gtrsim (\Delta V/2)^{-1}$, respectively.

\paragraph*{\textbf{Sharp Interface in a Neutron Star.---}} 
A longstanding open question is whether NS cores contain a 1st-order phase transition separating an exotic interior ({\it e.g.} quark matter) and a hadronic exterior \cite{Fukushima:2010bq,Rajagopal:1999cp,Page:2006ud,Alford:2013aca,Annala:2019puf}. If such a 1st-order transition exists, and if the surface tension between these phases is sufficiently large to disfavor mixed phases, a density/composition jump may develop on the phase boundary \cite{Baym:2017whm,Alvarez-Castillo:2017qki,2021Univ....7..493L,Buballa:2003qv}. Establishing or ruling out such a sharp transition would provide a valuable insight on NS physics and baryon-dense QCD, a regime where first-principle lattice calculations are prohibitive due to the fermion sign problem \cite{Nagata:2021ugx}.

Motivated by these possibilities, we crudely model an NS as a piecewise-uniform ball of mass $M$ and radius $R$, containing a density/composition jump at radius $R_{\rm jump}$ over a transition thickness $l$. We assume the neutrino matter potential is $V\approx-V_{\rm in}$ in the inner region ($r< R_{\rm jump}$) and $V\approx -V_{\rm in}+\Delta V$ in the outer region ($R_{\rm jump}<r<R$). In the main part of our analysis, we adopt the benchmark parameters $M=1.4M_\odot$, $R=10\text{ km}$,  $V_{\rm in}=33\eV$, and
\begin{align}\label{eq:benchmarks}
    \Delta V=20\eV,\quad l\ll \Delta V^{-1},\quad R_{\rm jump}=8\text{ km}.
\end{align}
These values are intended to be representative of the magnitude and location of the sharp interface considered in the literature, namely an $\mathcal{O}(1)$ change at $\mathcal{O}(1)$ of the NS radius; see {\it e.g.}~\cite{Pereira:2017rmp,Tonetto:2020bie,Han:2018mtj,Alvarez-Castillo:2017qki,Benic:2014jia,2016EPJA...52..232A,Ranea-Sandoval:2015ldr,Baym:2017whm}. The bulk of an NS is expected to become isothermal within a few years after birth due to efficient thermal conduction. We can treat the NS as an isothermal core with temperature $T_{\rm core}$ covered by a thin ($\sim 100\text{ m}$) envelope in which the temperature drops from $T_{\rm core}$ to the surface temperature $T_{\rm sur}$. We assume a non-accreting NS  and relate $T_{\rm core}$ and $T_{\rm sur}$ using the envelope fit from Refs.~\cite{1982ApJ...259L..19G,1983ApJ...272..286G} 
\begin{align}\label{eq:envelope}
     T_{\rm core}\approx \text{max}\left[100\eV \left(\frac{T_{\rm sur}}{8\times 10^4\text{ K}}\right)^{1.8},T_{\rm sur}\right].
 \end{align}
Here, we have fixed the surface gravity to $g_{\rm sur}=2\times 10^{12}\text{ m}/\text{s}^2$, consistent with our benchmark $M$ and $R$. The $\text{max}\left[\ldots\right]$ ensures $T_{\rm core}\geq T_{\rm sur}$ in general and $T_{\rm core}\approx T_{\rm sur}$ for $T_{\rm sur} \lesssim 4000\text{ K}$, as expected \cite{Yakovlev:2004iq}.

In models with a 1st-order phase transition, the scale length of the jump is typically set by the QCD scale, $l\sim \Lambda_{\rm QCD}^{-1}\sim \text{fm}$ \cite{Pereira:2017rmp,Tonetto:2020bie,Han:2018mtj,Alvarez-Castillo:2017qki,Benic:2014jia,2016EPJA...52..232A,Ranea-Sandoval:2015ldr,Baym:2017whm}. Importantly, the ``jump" need not be a true discontinuity. A smooth but sufficiently rapid crossover could still lead to a significant pair-production rate. In fact, the maximal, sharp-limit production rate is recovered as long as the characteristic width $l$ of the transition region satisfies $l\lesssim \Delta V^{-1}\sim 10\text{ nm}$; see Fig.~\ref{fig:lumi}. 
Using the sharp limit of Eq.~\eqref{eq:rateperareasharp} and integrating over the jump-interface area gives a neutrino production rate of \footnote{Neutrino-antineutrino pair creation can also be stimulated by neutrinos(antineutrinos) approaching the jump interface from the inside(outside), a process responsible for the Klein paradox \cite{1999PhR...315...41D}. For simplicity, we conservatively neglect this contribution. When the bound neutrinos are degenerate, we expect this contribution to be $\mathcal{O}(1)$ of the spontaneous production rate we consider.}
\begin{align}\label{eq:gradrate}
    \dot{N}_\nu^{\rm grad} \approx
   4\times 10^{38}\text{ s}^{-1} \left( \frac{N_f}{3} \right)
\left( \frac{\Delta V}{20\eV} \right)^3
\left( \frac{R_{\rm jump}}{8\text{ km}} \right)^2,
\end{align}
where $N_f$ is the number of neutrino flavors we consider. Gradient-produced neutrinos are pulled inward and become bound within $R_{\rm jump}$, while antineutrinos are repelled. In the absence of neutrino interactions other than forward scattering, the bound-neutrino population builds up until their total number reaches $N_\nu^{\rm deg}=N_f\Delta V^3/6\pi^2\times 4\pi R_{\rm jump}^3/3$, {\it i.e.},
\begin{align}\label{eq:Ndeg}
N_{\nu}^{\rm deg} \approx 2\times 10^{35}\left(\frac{\Delta V}{20\eV}\right)^3\left(\frac{R_{\rm jump}}{8\text{ km}}\right)^3,
\end{align}
whereupon they become degenerate, Pauli blocking further production~\cite{1990PhRvL..64..115L,Kachelriess:1997cr}. Such degenerate neutrinos backreact negligibly to the NS \footnote{At the rate in Eq.~\eqref{eq:gradrate}, bound neutrino states would fill in about $N_\nu^{\rm deg}/\dot{N}_\nu^{\rm grad}\sim 1\text{ day}$. Since the total mass of the degenerate neutrinos $\sim \Delta VN_\nu^{\rm deg}\sim 10\text{ kg}\ll M_\odot$ is tiny, they backreact negligibly to the NS.}.

 \begin{figure}[!t]
     \centering
     \includegraphics[width=\linewidth]{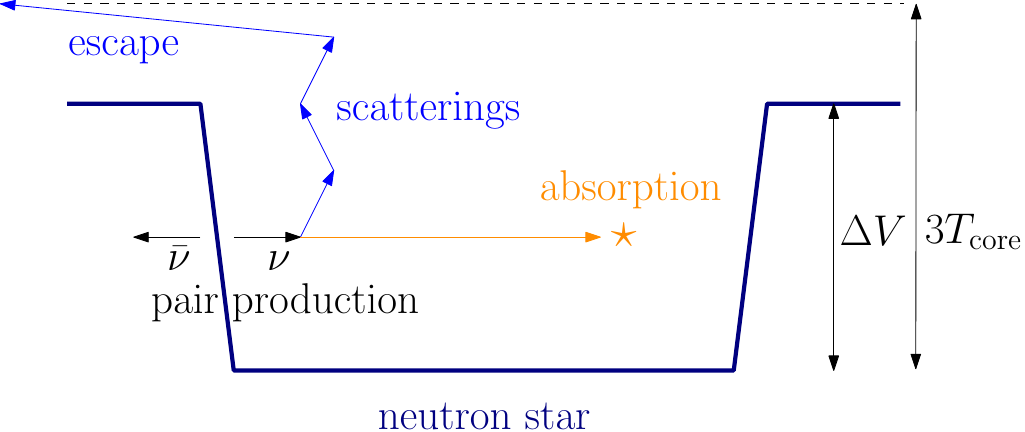}
     \caption{Illustration of pair production of bound neutrinos by a matter-potential jump in NS, their escape through upscattering when $3T_{\rm core}\gtrsim \Delta V$, and their absorption.}
     \label{fig:cloud}
 \end{figure}
 
\paragraph*{\textbf{Neutrino Absorption and Scattering.---}}
Bound neutrinos can be removed from the low-energy phase space by absorption on the NS matter and/or upscattering to energies above the matter-potential well allowing for escape, as sketched in Fig.~\ref{fig:cloud}. Both processes vacate the bound phase space and thus prevent gradient production from shutting off completely due to Pauli blocking. The evolution of the number of matter-potential-bound neutrinos $N_\nu$ is roughly captured by
\begin{align}
    \dot{N}_\nu=\dot{N}_\nu^{\rm grad}F-\frac{N_\nu}{\lambda_{\nu,\rm mfp}},
\end{align}
where $F$ accounts for Pauli blocking and $\lambda_{\nu,\rm mfp}$ is an effective mean free path for the processes that deplete the bound population, \textit{i.e.},~absorption or scattering. Assuming for simplicity that $F\sim1-N_\nu/N_\nu^{\rm deg}$, we find that the steady-state ($\dot{N}_\nu=0$) gradient-production rate is
\begin{align}\label{eq:NdotgradF}
    \dot{N}_\nu^{\rm grad}F_{\rm steady}\sim \frac{\dot{N}_\nu^{\rm grad} }{1+\lambda_{\nu,\rm mfp}\dot{N}_{\nu}^{\rm grad}/N_\nu^{\rm deg}},
\end{align}
which is maximal when $\lambda_{\nu,\rm mfp}\lesssim N_\nu^{\rm deg}/\dot{N}_\nu^{\rm grad}$ and reduces to  $N_\nu^{\rm deg}/\lambda_{\rm mfp}$ when $\lambda_{\nu,\rm mfp}\gg N_\nu^{\rm deg}/\dot{N}_\nu^{\rm grad}$. Note that $N_\nu^{\rm deg}/\dot{N}_\nu^{\rm grad}\approx 10R_{\rm jump}$ if Eq.~\eqref{eq:gradrate} applies. 

Beyond enabling continued gradient production, neutrino absorption and scattering can alter the thermal evolution of an NS dramatically. Below we specialize to a deconfined strange quark matter core \cite{Zdunik:2012dj,Lastowiecki:2011hh,Alford:2004pf}, although we stress that the broad-brush phenomena we discuss here are not limited to this particular phase and can arise analogously in other NS phases.

\begin{figure}[t!]
    \centering
    \includegraphics[width=\linewidth]{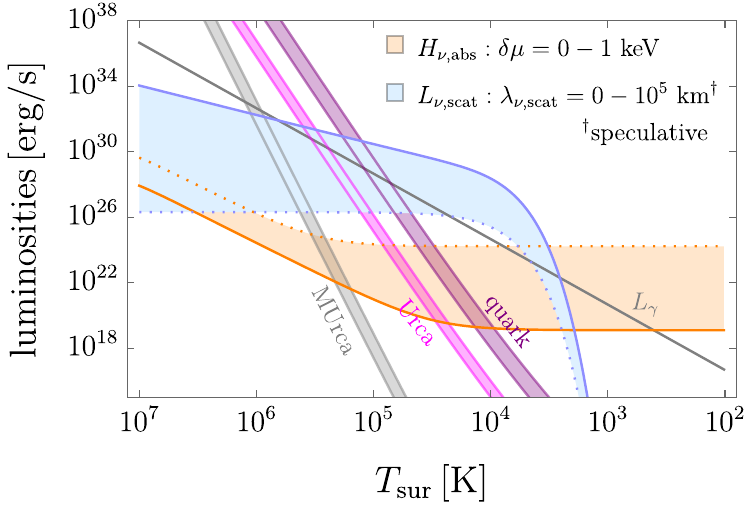}
    \caption{Luminosities as a function of surface temperature $T_{\rm sur}$. The orange band is the heating luminosity of absorbed gradient-produced neutrinos, given by~Eq.~\eqref{eq:Hnuabs}, with a chemical-potential imbalance varied from $\delta\mu=0$ (solid orange) to $\delta\mu=1\keV$ (dotted orange). The blue band is the escaping luminosity of upscattered gradient-produced neutrinos, given by Eq.~\eqref{eq:Lnugrad}, for scattering mean free paths between $\ll R_{\rm jump}$ (solid blue) and $10^{5}\text{ km}$ (dotted blue). The gray line is the thermal photon luminosity from the NS surface. The gray, magenta, and purple bands are the thermal-neutrino luminosities for MUrca, Urca, and quark processes, respectively, given in the Supplemental Material.}
    \label{fig:LvsTsur}
\end{figure}

\begin{figure}[t!]
    \centering
    \includegraphics[width=\linewidth]{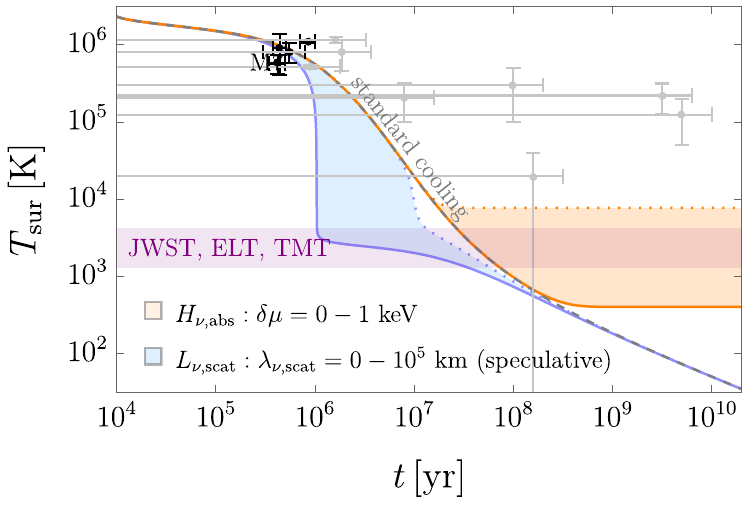}
    \caption{NS cooling curves: surface temperature $T_{\rm sur}$ as a function of age $t$, computed with Eq.~\eqref{eq:heateq}. The standard cooling curve (dashed-gray) accounts for the surface-photon emission $L_\gamma$ and the MUrca thermal-neutrino emission $L_{\nu,\rm th}$. The orange band additionally includes gradient-produced neutrino absorption heating $H_{\nu,\rm abs}$ for chemical-potential imbalance ranging from $\delta\mu=0$ (solid orange) to $\delta\mu=1\keV$ (dotted orange). The blue band additionally includes gradient-produced scattering cooling $L_{\nu,\rm scat}$ for scattering mean free paths ranging from $\ll R_{\rm jump}$ (solid blue) to $10^5\text{ km}$ (dotted blue). Here the $L_{\nu,\rm scat}$ and $H_{\nu,\rm abs}$ are considered in isolation, not simultaneously. These cooling and heating rates are plotted in Fig.~\ref{fig:LvsTsur}. The error-barred data points are the Magnificent Seven (M7) NSs from Ref.~\cite{Potekhin:2020ttj} and five oldest NSs from Ref.~\cite{Yanagi:2019vrr}. Those with kinematic age available are shown in black; those with only spindown age in light gray. The light-purple band shows the $T_{\rm sur}$ sensitivity of current and planned infrared telescopes, JWST, ELT, and TMT. }
    \label{fig:coolingcurve}
\end{figure}

In strange quark matter, electron neutrinos can be absorbed via $\nu_e s\rightarrow u e^-$. The mean free path for $E_\nu\sim \Delta V$ is $\lambda_{\nu,\rm abs}\sim 0.5\text{ m } \mu_u\mu_e/(\Delta V+T_{\rm core})^2$, assuming $s,u,e^-$ are in a chemical-potential equilibrium, $\mu_s=\mu_u+\mu_e+Q_{\rm vac}$, with $Q_{\rm vac}=m_s-m_u-m_e$ and $\mu_u\sim \mu_e\sim 100\MeV$ \cite{1982AnPhy.141....1I}. In this case, each absorption imparts an energy $\sim \Delta V$ on $u$ and $e^-$, which promptly thermalize via strong and electromagnetic interactions, heating the NS matter. However, small density changes ({\it e.g.}~from rotochemical evolution \cite{Reisenegger:1994be} or mechanical waves \cite{2000A&A...357.1157H}) as tiny as $\delta \rho/\rho\gtrsim 10^{-7}$ can generate a chemical imbalance $\delta\mu=\mu_s-\mu_u-\mu_e-Q_{\rm vac}$ in excess of $E_\nu\sim 10\eV$ \footnote{The chemical potential $\mu$ and density $n$ of an asymmetric species are related approximately as $\mu\sim n^{1/3}$.}, which can persist in old and cold NS where beta equilibration is slow. A nonzero $\delta\mu$ reduces Pauli blocking for $u$ and $e^-$, shortening the absorption mean free path to
\begin{align}
    \lambda_{\nu,\rm abs}\sim 0.5\text{ m}\frac{\mu_u\mu_e}{(\Delta V+T_{\rm core}+\delta\mu)^2}.
\end{align}
At the same time the chemical imbalance increases the energy released per absorption by $\delta\mu$. Both effects can enhance the heating rate by absorption substantially.

If the NS core is sufficiently hot that $3T_{\rm core}\gtrsim \Delta V$, the bound neutrinos can be upscattered above the matter-potential well and escape. In this case, both neutrinos and antineutrinos extract thermal energy from the medium and carry them away, thus cooling the core. When $\lambda_{\nu,\rm scat}\gtrsim R_{\rm jump}$, (anti)neutrinos escape after a few scatterings with energies $E_{\nu,\rm esc}\sim V_{\rm in}$. If $\lambda_{\nu,\rm scat}\ll R_{\rm jump}$ instead, (anti)neutrinos are temporarily trapped and diffuse, leaking out after many scatterings with energies close to $3T_{\rm core}$. A crude baseline estimate for the neutrino scattering mean free path in a homogeneous quark matter yields values in the ballpark of $\lambda_{\nu,\rm scat}\sim 2\times 10^{13}\text{ km}(T_{\rm core}/10\keV)^{-1}$, which implies an extremely suppressed steady-state gradient production, as per Eq.~\eqref{eq:NdotgradF}. However, gradient-produced neutrinos have long wavelengths of order $\Delta V^{-1}\sim 10\text{ nm}$, making their scattering highly sensitive to long-range correlations in the medium, which may reduce the $\lambda_{\nu,\rm scat}$ by orders of magnitude. Past studies have focused on correlations at scales $\lesssim (10\MeV)^{-1}\sim 10\text{ fm}$, while correlations at $\gg 10\text{ fm}$ scales remain unexplored. We therefore treat $\lambda_{\nu,\rm scat}$ as a free parameter in this exploratory study. Since we have not identified a concrete origin of the neutrino-scattering enhancement, our neutrino-absorption results should be considered more robust than our neutrino-scattering results.

 The heating rate from absorption $H_{\nu,\rm abs}$ and the cooling rate from scattering $L_{\nu,\rm scat}$ can be estimated as
\begin{align}
     L_{\nu,\rm scat}&\sim 2 \dot{N}_\nu^{\rm grad}F_{\rm steady} e^{-V_{\rm in}/3T_{\rm core}}\times E_{\nu,\rm esc}\label{eq:Lnugrad},\\
     H_{\nu,\rm abs}&~\sim \frac{1}{3} \dot{N}_\nu^{\rm grad}F_{\rm steady}\times (\Delta V+\delta\mu),\label{eq:Hnuabs}
 \end{align}
where the factor of 2 in $L_{\nu,\rm scat}$ accounts for neutrinos and antineutrinos while the Boltzmann factor $e^{-V_{\rm in}/3T_{\rm core}}$ captures the suppression of neutrino evaporation when $3T_{\rm core}\lesssim V_{\rm in}$. In $H_{\nu,\rm abs}$, the factor $1/3$ reflects that only $\nu_e$ (but not $\nu_{\mu,\tau}$ and antineutrinos) can be absorbed efficiently in strange quark matter. We take $E_{\nu,\rm esc}\sim V_{\rm in}+3T_{\rm core}e^{-\lambda_{\nu,\rm scat}/R_{\rm jump}}$ as a smooth interpolation between the $\lambda_{\nu,\rm scat}\gtrsim R_{\rm jump}$ and $\lambda_{\nu,\rm scat}\ll R_{\rm jump}$ cases. More details are given in the Supplemental Material Sec.~\ref{s:depletion}.

\paragraph*{\textbf{Neutron Star Cooling Curve.---}}  In the minimal cooling model \cite{Page:2004fy}, NSs older than $10^5\text{ yr}$ cool predominantly via thermal photon emission from the surface. Standard thermal neutrino luminosities from the core typically scale steeply with the core temperature as $T_{\rm core}^8$ or $T_{\rm core}^6$ \cite{Yakovlev:2004iq}. Consequently, for old NSs with $T_{\rm core}\ll \MeV$, these thermal neutrino emissions rapidly become negligible. By contrast, the heating amd cooling associated with gradient-produced neutrinos depend for the most part only mildly with $T_{\rm core}$. Their effects are therefore particularly important for old and cold NSs.

We model the thermal evolution of an NS as \cite{Yakovlev:2004iq,Page:2004fy}\footnote{For clarity, we neglect general-relativistic redshift factors, which only change luminosities by $\mathcal{O}(10\%)$.}
\begin{align}\label{eq:heateq}
    \dot{U}_{\rm NS}\approx &-L_\gamma-L_{\nu,\rm th}-L_{\nu,\rm scat}+H_{\nu,\rm abs}
\end{align}
where $U_{\rm NS}\approx 4\times 10^{41}\text{erg/s}\times(T_{\rm core}/100\eV)^2$ is the heat capacity of the NS for our assumed parameters, specified above Eq.~\eqref{eq:benchmarks}, $L_\gamma=(\pi^2/60)T_{\rm sur}^4\times 4\pi R^2$ is the thermal photon luminosity, $L_{\nu,\rm th}$ is a model-dependent thermal-neutrino luminosity, and $L_{\nu,\rm scat}$ and $H_{\nu, \rm abs}$ are, respectively, the scattering-induced cooling and absorption-induced heating rates from gradient-produced neutrinos. Although our focus is on regimes where $L_{\nu,\rm th}$ is subdominant, we include it for comparison using three representative thermal processes that bracket common NS models (see Supplemental Material for details \cite{Yakovlev:2004iq}), namely (1) $n\rightarrow pe\bar{\nu},\,pe\rightarrow n\nu$ (Urca), (2) $nN\rightarrow pNe\bar{\nu},\, pNe\rightarrow nN\nu$ (MUrca), and (3) $d\rightarrow ue\bar{\nu},\, ue\rightarrow d\nu$ (quark).  Fig.~\ref{fig:LvsTsur} shows that $L_{\nu,\rm scat}$ and $H_{\nu,\rm abs}$ can exceed $L_{\gamma}$ and $L_{\nu,\rm th}$ in certain $T_{\rm sur}$ ranges. 

In Fig.~\ref{fig:coolingcurve}, we plot our predicted cooling curves (surface temperature $T_{\rm sur}$ vs age $t$). We compare the standard-cooling benchmark where only $L_\gamma$ and MUrca $L_{\nu\bar{\nu}}^{\rm th}$ are included (dashed-gray) against scenarios that additionally include gradient-produced neutrinos. We start the evolution at $t=10^{4}\text{ yr}$ with $T_{\rm sur}\sim 100\eV$ and, for clarity, treat absorption heating (orange) and scattering cooling (blue) separately. For absorption, we calculate $H_{\nu,\rm abs}$ with Eq.~\eqref{eq:Hnuabs}, varying the chemical-potential imbalance from $\delta\mu=0$ (solid-orange) to $\delta\mu=1\text{ keV}$ (dotted-orange). The absorption-heating curves track the standard cooling at early times, but eventually freeze out to a constant $T_{\rm sur}$ set by the balance $H_{\nu,\rm abs}=L_{\gamma}$, analogous to the late-time plateaus found in rotochemical-heating \cite{Reisenegger:1994be,Reisenegger:1996ir,Fernandez:2005cg} and DM-heating \cite{Baryakhtar:2017dbj,Joglekar:2020liw,Acevedo:2019agu} scenarios. For scattering, we calculate $L_{\nu,\rm scat}$ using Eq.~\eqref{eq:Lnugrad}, varying $\lambda_{\nu,\rm scat}$ from $\ll R_{\rm jump}$ (solid-blue) to $10^{5}\text{ km}$ (dotted-blue). The scattering-cooling curves deviate from the standard-cooling one once $L_{\nu,\rm scat}\gtrsim L_\gamma,L_{\nu,\rm th}$ and develop a characteristic knee at $T_{\rm sur}\sim 3000\text{ K}$, corresponding to the onset of exponential suppression of $L_{\nu,\rm scat}$ when $3T_{\rm core}\lesssim \Delta V$. Thereafter, photon cooling takes over and the curves asymptote back to the standard cooling curve. 

Overall, gradient-produced neutrinos can either slow down NS cooling via $H_{\nu, \rm abs}$ or speed it up via $L_{\nu,\rm scat}$ (provided enhanced scattering) relative to standard expectations, in an NS-model-dependent but predictive way. While we have separated absorption and scattering to highlight each effect, both processes should generically operate simultaneously. Nevertheless, as Fig.~\ref{fig:coolingcurve} shows, absorption heating and scattering cooling are significant for different NS ages, $t\gtrsim 10^{7}\text{ yr}$ for absorption and $t\sim 10^6-10^{7}\text{ yr}$ for scattering. Thus, their combined effects will resemble their separated effects.

Throughout our analysis we take $\Delta V= 20\eV$ and $l\ll \Delta V^{-1}$, {\it i.e.}~the sharp limit in which $\dot{N}_\nu^{\rm grad}$ is maximized for a given $\Delta V$. Our absorption-heating results, however, apply more generally. In regimes where absorption heating can be important, we typically have $\dot{N}_\nu^{\rm grad}\gg N_\nu^{\rm deg}/\lambda_{\nu,\rm abs}$, and so the effective gradient-production rate in Eq.~\eqref{eq:NdotgradF} saturates at the absorption-limited value $N_\nu^{\rm deg}/\lambda_{\nu,\rm abs}$, which is independent of $\dot{N}_\nu^{\rm grad}$. Consequently, for $\Delta V=20\eV$ and neutrino masses $\sim 0.1\eV$, all our predictions remain unchanged for thicker interfaces, for $l\lesssim 4\times 10^{-4}\text{ m}(\delta\mu/10\eV)^2$ and before chiral-mixing becomes important. For weaker jumps, $\Delta V\ll 20\eV$, the absorption-limited heating is reduced, but can remain observable at $T_{\rm sur}\gtrsim 1000\text{ K}$ if $\Delta V\gtrsim 10\meV(\delta\mu/10\eV)^{-2}$ (as realizable in the crust) and the lightest neutrino mass is sufficiently light. Even without any jump, smooth gradients guarantee some gradient production and can yield observable heating if the lightest neutrino is ultralight. Cold isolated NS could, in principle, set a lower bound on the neutrino mass, given a model for $\delta\mu$. We expand on these points in End Matter.

\paragraph*{\textbf{Observational Prospects.---}}

These results motivate searches for old NSs with reliable ages and well-measured surface temperatures $T_{\rm sur}$ \cite{Chatterjee:2022dhp,Bramante:2024ikc}. At present, most $T_{\rm sur}$ measurements come from UV and X-ray data, which select relatively young NSs with high $T_{\rm sur}$. We show in Fig.~\ref{fig:coolingcurve} the Magnificent Seven (M7) isolated NSs (Table 2 of Ref.~\cite{Potekhin:2020ttj}) and the five oldest NSs in Table 1 of Ref.~\cite{Yanagi:2019vrr}, including the coldest known NS, J2144, and the old but anomalously warm PSR J2124, J0437, J0108, and B0950. Many other NSs are either too young to appear in our plot, exhibit complicating features (e.g.~strong magnetic fields or hot spots), or lacking precise $T_{\rm sur}$.

The M7 are broadly consistent with all the cooling curves shown, and thus do not strongly discriminate between standard cooling and scenarios with gradient-produced neutrinos. By contrast, the handful of anomalously warm old NSs lie above our absorption-heating curves for $\delta\mu\lesssim \keV$, if their spindown ages are faithful proxies for their true ages. These anomalous NSs can be explained by unreliable spindown ages, them being non-isolated, or stronger gradient-produced neutrino absorption corresponding to $\delta\mu\gtrsim \keV$, which can be realized in some NS models, particularly those featuring an energy gap that permits sizeable chemical imbalances \cite{2010A&A...521A..77P}. More conventional explanations remain viable as well, including  accretion \cite{Wijnands:2004pg,Wijnands:2017jsc,Shternin:2007md, 2009ApJ...698.1020B,Rutledge:2001nb}, rotochemical heating \cite{Reisenegger:1994be,Reisenegger:1996ir,Fernandez:2005cg}, DM heating \cite{Baryakhtar:2017dbj,Joglekar:2020liw,Acevedo:2019agu}, and others  \cite{Vigano:2013lea,Pons:2008fd,2010A&A...522A..16G}. 

JWST and the upcoming TMT and ELT have the capabilities to detect nearby NSs with $T_{\rm sur}\sim 1300-4300\text{ K}$. This $T_{\rm sur}$ range, shown in light purple in Fig.~\ref{fig:coolingcurve}, is particularly revealing, because our mechanism predicts pronounced departures from standard cooling there: either a heating-supported plateau or a knee-like cooling feature. In parallel, new and ongoing radio surveys with FAST, SKA, and CHIME will expand the catalog of nearby pulsars, providing targets for dedicated IR follow-up \footnote{If characteristic features consistent with gradient-produced-neutrinos are identified in the cooling curve, the underlying hypothesis of a sharp interface in the NS interior can be further tested by other means. The same interface can be constrained by NS mass-radius measurements, by stability considerations against perturbations \cite{Alford:2013aca,Pereira:2017rmp,2021Univ....7..493L,1983A&A...126..121S,Lindblom:1998dp}, and by distinctive gravity-pulsation quasinormal modes (g-modes) potentially accessible to gravitational-wave observations \cite{Sotani:2001bb,Miniutti:2002bh,Orsaria:2019ftf,Flores:2013yqa,Ranea-Sandoval:2018bgu,Andersson:1997rn,Andersson:2021qdq,2011PhRvD..83b4014S}.}.

\paragraph*{\textbf{Discussion.---}}
We have studied Schwinger-like neutrino-antineutrino production inside NSs sourced by steep density/composition gradients, as can arise in, but not limited to, models with 1st-order phase transitions. We find that these gradient-produced (anti)neutrinos open additional heating and potentially cooling channels whose rates can exceed those of thermal-photon and thermal-neutrino emissions in old NSs. Our work suggests that surface-temperature measurements of isolated and old NSs, coupled with reliable age estimates, can probe the state of matter inside NSs, baryon-dense QCD, and neutrino properties at low energies. Current and upcoming radio and infrared telescopes will soon be able to make such observations, making our results timely. 

In this first study, our analysis is highly simplified. Important next steps include repeating the study with realistic NS models and phases, and careful treatments of neutrino transport considering flavor/chiral oscillations and perhaps magnetic fields. While we have focused on a benchmark NS featuring a matter-potential step in Eq.~\eqref{eq:benchmarks}, our absorption-heating results (summarized in Figs.~\ref{fig:LvsTsur} and \ref{fig:coolingcurve}) apply more broadly, as discussed in the End Matter. In particular, these results remain unchanged for $l$s that are orders of magnitude thicker, whereas for smaller $\Delta V$, the heating decreases quantitatively while retaining the same qualitative behavior. For sufficiently large $l$ and/or sufficiently small $\Delta V$, the results become sensitive to the mass of the lightest neutrino eigenstate. Moreover, even without a matter-potential step, the smooth density profile of an NS guarantees some gradient production. If ultralight neutrinos exist, this contribution may be observable under plausible chemical-potential assumptions. Thus, our mechanism could, in principle, be used to set a lower bound on the lightest neutrino mass.

More broadly, gradient pair production is not limited to gauge fields and matter potentials, but can be generalized further to other background fields. This opens new avenues of exploration beyond the specific NS application within the SM. For example, the same effect occurs in simple extensions involving new fermion $\chi$ coupled to an  SM current $j_{\rm SM}^\mu$ through an operator of the form $j_{\rm SM}^\mu \bar{\chi}\gamma_\mu(c_V-c_A\gamma_5)\chi$, which mimics weak four-fermion interaction. If the produced particle is bosonic, the process can be increasingly Bose-stimulated in analogy with black hole superradiance \cite{Arvanitaki:2009fg,Arvanitaki:2024taq}, potentially increasing its detectability. A BSM field may itself play the role of the background field, \textit{e.g.}, an ultralight (psudo)scalar $\phi$ with derivative couplings to some SM fermions or a macroscopic dark matter state with appropriate interactions \cite{Ebadi:2021cte,Fedderke:2024hfy,Kaplan:2024dsn} could in principle gradient-produce SM particles. A systematic exploration of such possibilities and their phenomenology in compact objects, other extreme environments, or in the laboratory is an interesting direction for future work \cite{FutureWork}. 

Finally, the low-energy neutrino absorption and scattering processes considered here may also be relevant for the relic cosmic neutrino background (C$\nu$B). It would be worthwhile to revisit NSs as potential C$\nu$B probes \cite{Das:2024thc,Chauhan:2024deu,1983NCimL..38..174D} in light of these effects. We also note that gradient-produced (anti)neutrinos that escape the NS would contribute to the local (anti)neutrino spectrum at energies $\gtrsim 10\eV$, potentially with some asymmetry; see the Supplemental Material for more details.

\paragraph*{\textbf{Acknowledgments.---}}We thank Peter Denton, Savas Dimopoulos, Alex Friedland, Marios Galanis, Wick Haxton, Anson Hook, David E. Kaplan, Jim Lattimer, Madappa Prakash, Yong-Zhong Qian, and Ken Van Tilburg for valuable discussions. E.H.T.~is supported by NSF Grant PHY-2310429, Simons Investigator Award~No.~824870, the Gordon and Betty Moore Foundation Grant GBMF7946, and the U.S.~Department of Energy (DOE), Office of Science, National Quantum Information Science Research Centers, Superconducting Quantum Materials and Systems Center (SQMS) under contract No.~DEAC02-07CH11359. Y.W.~is supported by the Simons Investigator Grant No.~144924. This work was performed in part at the Aspen Center for Physics, which is supported by National Science Foundation grant PHY-2210452.

\bibliography{references}
\clearpage

\section*{End Matter}

\section{Neutron Star Density Gradients}

Steep density gradients may also arise in the absence of a first-order phase transision. The NS crust can host abrupt density/composition changes with radius, \textit{e.g.}, as a result of accretion and/or nuclear burning. A discontinuous change in the atomic number $Z$ and mass number $A$ at fixed pressure, $(A,Z)\rightarrow (A',Z')$, results in a baryon density jump $\Delta n_B/n_B\approx (Z/Z')(A'/A)-1 $  \cite{Chamel:2008ca,Miniutti:2002bh,1990MNRAS.245..508M,1987MNRAS.227..265F}. Separately, it has been suggested that in configurations where charged neutrality is enforced only globally, rather than locally, a density discontinuity can be maintained at the core-crust interface \cite{Rotondo:2011rj,Belvedere:2012uc,Voskresensky:2002hu}. More speculatively, BSM physics can also support sharp gradients. For example, lighter QCD axions may allow domains in which the $\theta$ angle varies from $0$ to $\pi$, which could result in abrupt changes to the density and composition \cite{Hook:2017psm,Kumamoto:2024wjd}.

\subsection{Robustness of absorption-heating results}
In absorption-heating scenarios, the observationally relevant absorption mean free path is typically very large. A representative estimate in chemically imbalanced matter is $\lambda_{\nu,\rm abs}\gtrsim 5\times 10^{13}\text{ m}(\delta\mu/10\eV)^{-2}$, appropriate in the regime $E_\nu+T_{\rm core}\lesssim \delta\mu$. For such a long $\lambda_{\nu,\rm abs}$, the steady-state pair-production rate is often limited not by $\dot{N}_\nu^{\rm grad}$ but by how quickly bound neutrinos are removed by absorption. When $\dot{N}_\nu^{\rm grad}\gtrsim N_\nu^{\rm deg}/\lambda_{\nu,\rm abs}$, Eq.~\eqref{eq:NdotgradF} implies that the effective production rate saturates at the absorption-limited rate $\dot{N}_\nu^{\rm grad}F\sim N_\nu^{\rm deg}/\lambda_{\nu,\rm abs}$, which is independent of $\dot{N}_\nu^{\rm grad}$. This means our absorption-heating results shown in Figs.~\ref{fig:LvsTsur} and \ref{fig:coolingcurve}, which assumes the $\dot{N}_\nu^{\rm grad}$ is given by the maximal rate in Eq.~\eqref{eq:gradrate}, have some degree of robustness under changes to the properties of the matter-potential responsible for gradient production. These results remain applicable for smaller $\dot{N}_\nu^{\rm grad}$ as long as 
\begin{align}\label{eq:absorptionlimited}
    \dot{N}_\nu^{\rm grad}\gtrsim \frac{N_\nu^{\rm deg}}{\lambda_{\nu,\rm abs}^{\rm min}}\sim 10^{30}\text{ s}^{-1}\left(\frac{\delta\mu}{10\eV}\right)^2\left(\frac{N_\nu^{\rm deg}}{2\times 10^{35}}\right).
\end{align}

A strong first-order transition with large surface tension can produce a sharp interface with $l\ll \Delta V^{-1}$. Weaker surface tension or mixed phases can increase $l$ to the point that the gradient-production transitions to the gradual regime $l\gtrsim \Delta V^{-1}$, where $\dot{N}_\nu^{\rm grad}\sim \dot{n}R_{\rm jump}^2l\propto \Delta V^2/l$, provided there are sufficiently light neutrino mass eigenstates. Nevertheless, as long as Eq.~\eqref{eq:absorptionlimited} holds, the pair-production stays in the absorption-limited regime and the heating rates are unchanged. This makes our absorption-heating signatures robust to large theoretical uncertainty in $l$. At fixed $\Delta V=20\eV$, $l$ can be increased up to macroscopic values $\sim 1\text{ m}(\delta\mu/10\eV)^2$ without violating Eq.~\eqref{eq:absorptionlimited}, provided at least one neutrino species is light enough to avoid exponential suppression of the pair-production rate and chiral mixing. For neutrino masses $m_\nu\sim 0.1\eV$, our absorption-heating results hold for $\Delta V=20\eV$ and at most until $l\lesssim 4\times 10^{-4}\text{ m}(\delta\mu/10\eV)^2$ but before chiral mixing becomes important. 

\subsection{Smaller matter-potential step}
If $\Delta V\ll 20\eV$, both the bare pair-production rate $\dot{N}_\nu^{\rm grad}$ and the fillable bound phase space $N_\nu^{\rm deg}$  decrease. In this case, it is important to distinguish between the jump $\Delta V$ which controls production at the jump interface and the full well depth $V_{\rm well}\sim 10-30\eV$ set by the smooth density profile, which controls the typical energy of bound neutrinos once they move away from the jump interface. The jump produces neutrinos with radial and transverse momenta $p_r, p_\perp\lesssim \Delta V$ near the jump interface. These neutrinos initially occupy a narrow band in energy $E=|\bm{p}|+V(r)\in[-\Delta V,0]$ and angular momentum $|L|=p_\perp r\lesssim \Delta V R$. Since the matter-potential is static and approximately spherically symmetric, both $E$ and $L$ are conserved. The phase-space integral for $N_\nu^{\rm deg}$ can be evaluated in terms of $E$ and $L$ as $N_\nu^{\rm deg}\sim \int d^3xd^3p=R^3\int [(E-V)/(r^2p_r)] dE LdL\sim R^3\Delta V^3$.  Hence, the absorption-limited condition Eq.~\eqref{eq:absorptionlimited} remains unchanged and the resulting rate scales as $\dot{N}_\nu^{\rm grad}F\propto \Delta V^3$. The absorption plateau in NS cooling curve remains potentially observable at $T_{\rm sur}\gtrsim 1000\text{ K}$ as long as $\Delta V\gtrsim \Delta V_{\rm obs}=10\meV(\delta\mu/10\eV)^{-1}$. 

A small potential step satisfying $\Delta V\gtrsim  \Delta V_{\rm obs}$ with a sufficiently sharp interface $l\ll \Delta V^{-1}$ may naturally occur in the NS crust where composition changes can be abrupt. This possibility is interesting because NS crusts are relatively well understood and such transitions may be common.  At such small $\Delta V$, the production may become sensitive to finite neutrino mass effects, including the chiral mixing which we have neglected in our production rate calculations. For simplicity, we assume here that the lightest neutrino mass eigenstate is small enough that chiral mixing remains negligible. Furthermore, we note that neutrino absorption in this crustal discontinuity scenario may occur via nuclear channels such as $\nu_en\rightarrow pe^-$ in addition to the quark process $\nu_es\rightarrow ue^-$ if quark matter exists deeper in.

\subsection{Guaranteed production from smooth gradients}

Even in the absence of a density jump, an NS has a smoothly varying potential $V(r)$ set by its density and composition profile. This provides an irreducible floor for gradient production, which can be estimated using the gradual-limit rate applied locally and integrated over the NS. Such gradient production was considered in \cite{loeb1990bound,kachelriess1998neutrino}. Taking a characteristic gradient $g\sim \Delta V/R$ with $\Delta V=33\eV$ and $R\sim 10\text{ km}$, the total gradient-production rate is $\dot{N}_\nu^{\rm grad}\sim g^2R^3\sim 10^{29}\text{ s}^{-1}$, assuming the lightest neutrino is light enough to avoid exponential suppression and significant chiral mixing. In this case, Pauli blocking is completely avoided provided $\lambda_{\nu,\rm abs}\lesssim 10^{15}\text{ m}$. In the absence of Pauli blocking, the resulting heating rate for $E_\nu+T_{\rm core}\lesssim \delta\mu$ is $H_{\nu,\rm abs}\sim 10^{20}\text{ erg/s}(\delta\mu/\keV)$, which can be comparable to the surface-photon emission at $T_{\rm sur}=10^3\text{ K}$, $L_\gamma\sim 10^{20}\text{ erg/s}$. Thus, if an isolated NS is observed to cool below $T_{\rm sur}\sim 10^{3}\text{ K}$ without evidence for an internal heating mechanism, one could in principle place a lower bound on the lightest neutrino mass. The robustness of this bound relies on a sufficient understanding of the evolution of the chemical imbalance $\delta \mu$, e.g., from rotochemical effects. 

Any additional sharpening of $V(r)$ due to a spatial phase transition, a crustal composition jump, global-neutrality effect, etc would increase $g$ locally and would boost the gradient-production rate and absorption heating relative to the smooth contribution discussed above.

\section{Mean-Field Backreaction}

Energy conservation implies that the NS must supply the energy carried away by the produced (anti)neutrinos. In the original Schwinger process, the pair-produced charges partially screen the electric field background, reducing the electric-field energy density \cite{Akhmedov:2009vs,Kluger:1992gb,Bloch:1999eu}. An analogous backreaction occurs here: the energy carried by emitted neutrinos and antineutrinos is compensated by a decrease in the potential energy of the NS constituents, say, neutrons. Newly produced neutrino-antineutino pairs have a prompt energy of $\sim 2m_\nu$, but the neutrinos (antineutrinos) are subsequently accelerated inward (outward), gaining a typical kinetic energy of $\sim \Delta V/2\gg m_\nu$. Thus, producing the pairs costs $\sim \Delta V n_\nu$ of matter-potential work per unit volume. The produced pairs, in turn, reduce the matter potential felt by \textit{neutrons} locally, by $\Delta V_n\sim 1/4\times2\sqrt{2}G_Fn_\nu$, which in turn lowers their potential-energy density by $\sim \Delta V_n n_n=\Delta V n_\nu$ \cite{1995PhRvD..52.5459F}. This is what pays for the energy carried by gradient-produced neutrinos. Confirming energy conservation accounting for $\mathcal{O}(1)$ factors requires integrating over the emitted spectrum. We do not attempt this here and use the above argument as an order-of-magnitude consistency check.

In the case with efficient scattering, the reduction in the matter potential by the presence of gradient-produced (anti)neutrinos exist only as long as the (anti)neutrinos remain in the NS. The mean-field backreactions from a given (anti)neutrino is lifted as soon as its escapes the NS. By contrast, in the absorption case, a $\nu_e$ gets absorbed far more efficiently than a $\bar{\nu}_e$ does. As the NS continually absorbs $\nu_e$ and repels $\bar{\nu}_e$, it slowly develops a net increase in electron lepton number $L_e$ over time. For the typical $\lambda_{\nu,\rm abs}\gg R_{\rm jump}$, the cumulative change $\Delta L_e\sim \dot{N}_{\nu}^{\rm grad}Ft$ remains negligible compared to the total $L_e
 \sim Y_eN_B\sim 10^{55}$ of an NS. Nevertheless, this bookkeeping is important energetically: the small $\Delta L_e$ slightly lowers the matter-potential energy and is what  supplies the energy carried away by escaping $\bar{\nu}_e$. Eventually the matter potential inside and outside $R_{\rm jump}$ will be leveled when enough electron lepton number accumulates in the central region to cancel out baryon's contributions. However, for observationally relevant $\nu_e$ absorption mean free paths of $\lambda_{\nu,\rm abs}\gtrsim 5\times 10^{13}\text{ m}(\delta\mu/10\eV)^{-2}$, this takes a total time of $L_e/(\dot{N}_\nu^{\rm grad}F)\sim 2\times 10^8\text{ Gyr}(\delta\mu/10\eV)^{-2}$, which is typically many orders of magnitude longer than the current age of the universe.

\section{Majorana Neutrinos}

Gradient pair-production can also occur for Majorana neutrinos. Majorana neutrinos have a purely axial coupling, $c_V=0$ and $c_A=1$, corresponding to chiral couplings $-c_L=c_R=1$. As a result, produced pairs have opposite helicities. If one naively inserts these couplings into Eqs~\eqref{eq:rateperareasharp} and \eqref{eq:rateperareagradual} and sums over both chiral sectors as in the Dirac case, one finds production rates larger by a factor of two relative to the Dirac case. However, for a Majorana field, the two chiral components are not independent degrees of freedom (they are related by charge conjugation). Counting both chiral sectors would double-count the same physical states. Keeping only one independent chiral sector, the gradient production rates are the same for Dirac and Majorana neutrinos.

Neutrinos with observable effects are typically highly energetic. For such neutrinos, amplitudes for processes that distinguish Dirac from Majorana neutrinos are suppressed by $m_\nu/E_\nu\ll 1$. This difficulty in experimentally testing the Dirac vs Majorana nature of neutrinos is known as the ``Dirac-Majorana confusion theorem" \cite{1982PhLB..112..137K,Kayser:1982br}; see also Ref.~\cite{Bigaran:2025kod,Kim:2021dyj,Akhmedov:2024qpr}. Our scenario is unique in that gradient-produced neutrinos can be very low energy, $E_\nu\sim 10\eV$ (generically set by the full depth of the matter potential, not necessarily the step $\Delta V$), while still have observable consequences through absorption heating and/or scattering cooling in NS. For $m_\nu\sim 0.1\eV$, one has $m_\nu/E_\nu\sim 10^{-2}$, which is still ultrarelativistic but not astronomically small. Given the long timescales relevant for neutrino transport and depletion in NS \cite{Dobrynina:2016rwy}, even $\mathcal{O}(m_\nu/E_\nu)$ effects may be relevant. This makes it plausible, at least in principle, that a careful treatment of finite-mass effects could reveal different signatures for Dirac vs Majorana neutrinos in this setting.

\include{SupplementalMaterial}

\end{document}

%% file: SupplementalMaterial.tex
\widetext

\begin{center}
    \textbf{\large Supplemental Material}\\ \vspace{5pt}
    \textbf{Gradient-Produced Neutrinos}\\ 
    \vspace{3pt}
    Erwin H. Tanin and Yikun Wang
\end{center}
\setcounter{equation}{0}
\setcounter{figure}{0}
\setcounter{table}{0}
\setcounter{secnumdepth}{3}
\setcounter{page}{1}
\makeatletter
\renewcommand{\theequation}{S\arabic{equation}}
\renewcommand{\thefigure}{S\arabic{figure}}
\renewcommand{\bibnumfmt}[1]{[S#1]}
\renewcommand{\citenumfont}[1]{S#1}
\renewcommand\thesection{S\arabic{section}}
\renewcommand\thesubsection{S\arabic{section}.\arabic{subsection}}
\renewcommand\thesubsubsection{S\arabic{section}.\arabic{subsection}.\arabic{subsubsection}}

\section{Dispersion Relations in Various Limits}
\label{sec:apprate}

Consider a fermion $\psi$ interacting with an external vector background $j_{\rm ext}^\mu$ as in Eq.~\eqref{eq:Lag}. We specialize to a purely timelike background that depends only on one spatial coordinate $z$, $j_{\rm ext}^\mu=(V(z),\vec{0})$. The Lagrangian then reduces to
\begin{equation}
\mathcal{L} = \overline{\psi} (i \slashed{\partial} - m) \psi - V(z)\, \overline{\psi} \, \gamma^0 (c_V - c_A \gamma^5) \,\psi.
\end{equation}
The Dirac equation reads
\begin{equation}
\big[ i \slashed{\partial} - m - V(z)\, \gamma^0 (c_V - c_A \gamma^5) \big] \psi = 0,
\end{equation}
and the single-particle Hamiltonian reads
\begin{equation}
\hat{H} = m \gamma^0 + \bm{\alpha} \cdot \bm{{p}} + V(z) (c_V - c_A \gamma^5),
\end{equation}
where $ \alpha^i = \gamma^0\gamma^i$, and $\bm{{p}} = - i \nabla$.

For $m=0$, the Hamiltonian decouples into $2\times 2$ blocks in the chiral basis,
\begin{equation}
\hat{H} |_{m=0}  =
\begin{pmatrix}
- \bm{\sigma} \cdot \bm{{p}} + V (c_V + c_A) & 
0 \\
0  &
\bm{\sigma} \cdot \bm{{p}} + V (c_V - c_A)
\end{pmatrix}.
\end{equation}
 The block-diagonal structure of $\hat{H} |_{m=0}$ means that the left- and right-chiral Weyl components $\chi_{L,R}$ propagate independently. In a massless theory these chirality eigenstates are also helicity eigenstates which satisfy $ \bm{\sigma} \cdot \bm{\hat{p}} \,\chi_{L,R} = h_{}\chi_{L,R}$, which lead to the following dispersion relations for the left- and right-handed Weyl spinors 
\be\ba \label{eq:weyl-dis}
&E_L = - h |\bm{p}| + V (c_V + c_A),\\
&E_R = + h |\bm{p}| + V (c_V - c_A).
\ea \quad\quad\quad\quad\quad\text{(massless)}\ee
For left-handed Weyl sector, the $h = -1$ state is a left-handed particle, and the $h=+1$ state is a right-handed antiparticle, and similarly for the right-handed Weyl sector (note that $\overline{\chi_L} = \overline{\chi}_R$). 

When $m\neq0$, the chiral components mix and the Hamiltonian is no longer block diagonal. The dispersion relation can be found from the relation $0={\rm det} \,[ \hat{H} - E]$, which can be mathematically reduced to 
\be\ba \label{eq:det}
0&= {\rm det} \,[ (E - c_V V)^2 - c_A^2 V^2 - \bm{p}^2 + 2 c_A V \bm{\sigma} \cdot \bm{{p}} + i (c_V - c_A) \bm{\sigma} \cdot \bm{\nabla} V - m^2].
\ea \ee
We now take 4-spinor $\psi$ to be a helicity eigenstate, $\bm{S} \cdot \bm{{\hat{p}}}\, \psi = h \psi $, where $\bm{S}$ is the spin operator, which reduces to $\bm{\sigma} \cdot  \bm{{\hat{p}}} \,\chi_{L,R} = h \, \chi_{L,R}$, and allows substituting $\bm{\sigma} \cdot  \bm{p}\rightarrow h |\bm{p}|$ in Eq.~\eqref{eq:det}. In asymptotic regimes where the WKB approximation $\bm{\nabla} V/V \ll |\bm{p}|$ holds, the dispersion relations for these helicity eigenstates simplify as
\be\ba 
E_{\pm}=  c_V V \pm \sqrt{ (|\bm{p}| - c_A h V )^2   + m^2},
\ea \quad\quad\quad\quad\quad\text{(WKB)}\ee
where the branch with a positive (negative) frequency $E$ corresponds to particle (antiparticle). Taking the massless limit $m\rightarrow 0$ from here yields
\be\ba
E_{\pm} |_{m \to 0} = 
c_V V \pm \big| |\bm{p}| - c_A h V \big|.
\ea \quad\quad\quad\quad\quad\text{(WKB, massless)}\ee
The two $\pm$ branches, together with the two helicities $h=\pm 1$, reproduce the four massless branches in Eq.~\eqref{eq:weyl-dis} after identifying which $h_{}$ corresponds to $E_+$ versus $E_-$:
\begin{align}
    E_{L,R}= E_{\mp\,{\rm sgn}( h|\bm{p}| -  c_A V)} |_{m \to 0}.
\end{align}
In particular for $c_AV>|\bm{p}|$, the matching is as shown in Table~\ref{tb:id}. To avoid confusion, we note that the handedness naming in the ``Branch" column follows the standard convention and refers to helicity ($h=+1\rightarrow \text{right-handed}$ and $h=-1\rightarrow \text{left-handed}$) not chirality. 

In a spatially varying potential, the positive-frequency branch at one location can cross the negative-frequency branch at another location, enabling mixing between them. This is the origin of particle-antiparticle production in the presence of a potential gradient. For example, in Table~\ref{tb:id}, in the massless limit, a pair production creates a particle and an antiparticle from the same row block {\it e.g.}~$E_+(h=+1)$ and $E_+(h=-1)$. This is simply because the Dirac equation breaks into two decoupled parts in the massless limit, and modes in the same row block belong to the same Dirac-equation part. This is evident in the chiral picture ($E_{L,R}$), and provides a guideline for matching onto the branches of the (massive) Dirac spinor ($E_\pm$). Pair production can also occur in the case of a Majorana fermion. Majorana fermions have $c_V=0$ and they only reside in one chiral sector (either left {\it or} right), within which, two different helicity states can mix in the presence of a potential.

\begin{table}[h]
\centering
\renewcommand{\arraystretch}{1.8}
\begin{tabular}{ccccc}
\hline
Dirac & Weyl & $h$ & Branch & frequency\\
\hline
$E_+$ & $E_L$ & $+1$ & right-handed antiparticle & $-|\bm{p}|+ (c_V + c_A ) V$ \\
$E_+$ & $E_L$ & $-1$ & left-handed particle &$ |\bm{p}|+ (c_V +c_A) V $\\
\hline
$E_-$ & $E_R$ & $+1$ & right-handed particle & $ |\bm{p}|+ (c_V - c_A) V$ \\
$E_-$ & $E_R$ & $-1$ & left-handed antiparticle &$ -|\bm{p}|+ (c_V - c_A) V$\\
\hline
\end{tabular}
\caption{Positive and negative frequency branches for $c_A V > | \bm{p}|$. Note that $(\psi_L)^c = (\psi^c)_R$.}
\label{tb:id}
\end{table}

\section{Generalizing Pure-Vector Sauter-Potential Pair-Production Rate to Arbitrary Vector and Axial-Vector Couplings}

\label{sec:solve}
In this section, we show that, in the small mass limit, the pair-production calculation for general vector and axial-vector couplings $(c_V, c_A)$ is mathematically identical to the solved problem of pair-production of charged particle-antiparticle pairs in an electric field, which corresponds to a pure-vector coupling $(c_V=1,c_A=0)$. As a result, the known pair-production rate obtained in the pure-vector case can be trivially generalized to general vector and axial-vector couplings by trivial rescalings, after appropriate decompositions into chiral sectors.

Consider spinor modes with a definite frequency $E$ and definite transverse momentum $\bm{p}_\perp=(p_x,p_y)$. For such modes, the temporal and transverse-spatial dependence is trivial, $\psi \propto e^{- i (E t - \bm{p}_{\perp} \cdot \bm{x}_{\perp})}$, and only the $z$-dependence remains nontrivial. The time-independent Dirac equation of such modes read
\be\ba \label{eq:}
0= 
& \begin{pmatrix}
-m & 
E - c_V V +  \left[  c_A V + \bm{\sigma}_{\perp} \cdot \bm{p}_{\perp} - i \sigma_3 \partial_z \right] \\
E - c_V V - \left[ c_A V + \bm{\sigma}_{\perp} \cdot \bm{p}_{\perp} - i \sigma_3 \partial_z \right] &
-m
\end{pmatrix}\psi.
\ea \ee
Here, we use mostly-minus metric, the Weyl basis for the gamma matrices, and write $z \equiv x_{3}$. We would like to reduce this Dirac equation to Schrodinger-like scalar equations. Note that for a generic $z$-dependent potential $V$, this cannot be done by a simple helicity projection, because the helicity is not conserved (since $p_z$ is not conserved).  To reduce the equation, we use instead the following ansatz for $\psi$:
\begin{equation}\label{eq:DiracwithRHS}
\psi =  
e^{- i (E t - \bm{p}_{\perp} \cdot \bm{x}_{\perp})}
\Big[
E \gamma^0
+ \bm{\gamma}_{\perp} \cdot \bm{p}_{\perp} - i \gamma^3 \partial_z + m - V(z) \gamma^0 (c_V + c_A \gamma^5)  
\Big]
\begin{pmatrix}
\chi_1 \\ \chi_2
\end{pmatrix},
\end{equation}
where $\chi_{1,2}(z)$ are two-spinors. The Dirac equation then becomes second-order differential equations for $\chi_{1,2}$,
\be\ba \label{eq:Diracchi12}
& \Big[
(E - c_R V )^2  - \bm{p}_{\perp}^2 + \partial_z^2 - m^2
+ i c_R  V' \sigma_3
\Big]\chi_1  = - 2 m c_A V \chi_2, \\
& \Big[
(E - c_L V )^2 - \bm{p}_{\perp}^2 + \partial_z^2 - m^2
- i c_L  V' \sigma_3
\Big]\chi_2  = 2 m c_A V \chi_1 ,
\ea\ee
where $V'=\partial_zV$ and we have defined $c_L \equiv c_V+c_A$, and $c_R \equiv c_V-c_A$ for notational simplicity. Because the only remaining matrix structure is due to $\sigma_3$, the equations are diagonal in the eigenbasis of $\sigma_3$, and we expand $\chi_{1,2}$ as \begin{equation}
\chi_{\alpha,s} (z) = \varphi_{\alpha,s} (z)\, u_s, \quad {\rm for} \quad \alpha = 1,2, \quad s= \pm 1, 
\end{equation}
where $u_{+} = (1,0)^T$, $u_{-} = (0,1)^T$, $\sigma^3 u_s = s\, u_s$, and $s$ is the spin index. With this expansion, we have reduced the Dirac equation to a Schrodinger-like form in terms of $\varphi_{\alpha,s}$. However, the two sectors $\alpha=1,2$ remain coupled if $m\neq 0$.

Since our application focuses on ultrarelativistic neutrinos, we now adopt a controlled approximation to remove the $\alpha=1,2$ mixing. In the limit $m\ll E,|\bm{p}|,|c_AV|$, the right-hand sides of Eq.~\eqref{eq:Diracchi12} can be dropped to leading order. Physically, this corresponds to neglecting rare chirality flips, which are suppressed by $m/E$. A convenient set of basis modes for the left and right sectors are
\be\ba \label{eq:}
\psi_{L,s}\, \approx \mathcal{N}_{L} \, e^{- i (E t - \bm{p}_{\perp} \cdot \bm{x}_{\perp})} 
\Big[
(E - c_L V ) \gamma^0
+ \bm{\gamma}_{\perp} \cdot \bm{p}_{\perp} - i \gamma^3 \partial_z 
\Big]
\begin{pmatrix}
0 \\[1.0em]
u_s
\end{pmatrix} 
\varphi_{2,s}(z)
,\\
\psi_{R,s}\, \approx \mathcal{N}_{R} \, e^{- i (E t - \bm{p}_{\perp} \cdot \bm{x}_{\perp})} 
\Big[
(E - c_R V ) \gamma^0
+ \bm{\gamma}_{\perp} \cdot \bm{p}_{\perp} - i \gamma^3 \partial_z 
\Big]
\begin{pmatrix}
u_s \\[1.0em]
0
\end{pmatrix}  \varphi_{1,s}(z).
\ea\ee
Here $\mathcal{N}_{L,R}$ are normalization factors set by, {\it e.g.}, the incident flux in a scattering problem.  The Schrodinger-like Dirac equations for $\varphi_{\alpha,s}$ then decouple into two independent second-order differential equations
\be\ba 
\label{eq:reducedDirac}& \Big[
(E - c_R V )^2 - \bm{p}_{\perp}^2 + \partial_z^2 - m^2
+ i s \, c_R  V'
\Big]\varphi_{1,s} (z)  \approx 0, \\
& \Big[
(E - c_L V )^2 - \bm{p}_{\perp}^2 + \partial_z^2 - m^2
- i s \, c_L  V'
\Big]\varphi_{2,s} (z)  \approx 0.
\ea\ee
Here we keep the $m^2$ on the left-hand sides so that in regions where $V$ is approximately uniform, the above reproduce the WKB dispersion relations of the previous subsection. We keep this $m^2$ dependence for the sole purpose of making a connection to the known, exponentially suppressed production rate in the gradual limit, where the $m^2$ term is important. However, the terms in the r.h.s.~of~Eq.~\eqref{eq:DiracwithRHS}, which introduce chiral mixing, can dominate over this $m^2$ term. Thus, chiral mixing would introduce important corrections to our results for sufficiently large $l$. Notice that, for the sharp limit ($l \ll \Delta V^{-1}$) with $\Delta V\gg m$, within $-l \lesssim z \lesssim l$, where the pair production dominantly occurs, one has $V' \sim \Delta V/l \gg m \Delta V \gg m^2$, which suggests that chiral mixing, as well as the $m^2$ term, can be legitimately neglected in the pair-production calculation. Once the pairs are produced and become asymptotically free, chiral oscillation could happen as they freely propagate. We defer a careful study of the mass dependence of the rate for $c_A\neq 0$ for future work. Accordingly, we have removed all mass dependence in our analysis in the main text.

We are now ready to obtain the pair-production rate by the Sauter potential $V(z)=(\Delta V/2)\tanh(z/l)$ of Eq.~\eqref{eq:V}. The reduced Dirac equations that we found in Eq.~\eqref{eq:reducedDirac} are mathematically identical to the Dirac equations (similarly reduced to a Schrodinger-like form) of a Dirac fermion in a pure-vector potential whose temporal component is $V$. This can be seen upon a substitution $V\rightarrow c_LV(z)$ for the left sector and a substitution $V\rightarrow c_RV(z)$ together with a spin-index flip $s\rightarrow -s$ for the right sector. Therefore, the known results derived for a Dirac fermion in a pure-vector background $(c_V=1,c_A=0)$ \cite{Fedotov:2022ely,Gelis:2015kya,Chervyakov:2009bq,Chervyakov:2011nr,1999PhR...315...41D}   can be generalized to arbitrary $(c_V,c_A)$ via these substitutions. Furthermore, we can assume without loss of generality that $\Delta V\geq 0$. The problem with negative $\Delta V$ can be mapped into the problem with 
positive $\Delta V$ by the reflection $z\rightarrow -z$, which flips the propagation directions of the produced particle and antiparticle but does not change the total pair-production rate. Accordingly, the total production rate in the zero-mixing approximation is given by the sum of the left- and right-sector rates computed with positive-definite effective jumps $|c_{L,R}|\Delta V$.

The pair-production rate per unit transverse area can be written as
\be\ba \label{eq:ndotperp}
 \dot{n}_{\perp} = \dot{n}_{L,\perp} + \dot{n}_{R,\perp} = \sum_{i=L,R} \int \frac{d E d^2 \bm{p}_{\perp}}{(2\pi)^3} N_i ( {E,\bm{p}_{\perp}} ),
\ea\ee
where $N_i(E,\bm{p}_\perp)$ is the expectation value of the occupation number of produced particle-antiparticle pairs in mode $(E,\bm{p}_\perp)$ for sector $i=L,R$. This quantity has previously been calculated for the pure-vector $(c_V=1,c_A=0)$, Sauter potential case \cite{Fedotov:2022ely,Gelis:2015kya,Chervyakov:2009bq,Chervyakov:2011nr,1999PhR...315...41D}. Using the aforementioned substitutions we can generalize the $N_i(E,\bm{p}_\perp)$ to arbitrary $(c_V,c_A)$ as \begin{align}\label{eq:Ni}
  N_i (E,\bm{p}_{\perp})
  =  \frac{\sinh \left( \pi l p_{i,z}^{+} \right) \sinh \left(  \pi l  p_{i,z}^{-} \right)}{ \sinh \frac{\pi l}{2} \left(\Delta V_{i} +  p_{i,z}^{+} -  p_{i,z}^{-}  \right) \sinh \frac{\pi l}{2} \left( \Delta V_i - p_{i,z}^{+} +  p_{i,z}^{-}  \right)},
\end{align}
where $p_{i,z}^\pm$ are the asymptotic longitudinal ($z$-direction) momenta of the produced excitations at $z\rightarrow \pm\infty$
\be\ba\label{eq:momenta}
p_{i,z}^{\pm} \equiv \sqrt{ \left( p_{i,0}^{\pm} \right)^2 - |\bm{p}_{\perp}|^2 - m^2 },\quad {\rm with}
\quad p_{i,0}^{\pm} \equiv E \mp \frac{1}{2} \Delta V_i\quad {\rm and}\quad \Delta V_i\equiv |c_i| \Delta V. 
\ea\ee
Requiring the asymptotic longitudinal momenta $p_{i,z}^\pm$ be real imposes the condition $| p_{i,0}^{\pm} | =|E\mp \Delta V_i/2| \ge \sqrt{|\bm{p}_{\perp}| + m^2 }$. Pair-production occurs in the supercritical/Klein region, namely the range of $(E,\bm{p}_\perp)$ for which positive-frequency (particle) modes on one side of the potential connect to the negative-frequency (antiparticle) modes on the other. With our convention, the asymptotic particle(antiparticle) modes reside at  $z\rightarrow -\infty$($z\rightarrow+\infty$), \textit{i.e.}, $p_{i,0}^->0$ ($p_{i,0}^{+}<0$). These signs for $p_{i,0}^\pm$ and the reality condition on $p_{i,z}^\pm$ together imply
\be\ba\label{eq:phasespace}
\sqrt{|\bm{p}_{\perp}| + m^2}-\frac{1}{2} \Delta V_i 
 \leq E \le -\sqrt{|\bm{p}_{\perp}| + m^2}+\frac{1}{2}  \Delta V_i. \quad \quad \text{(klein/supercritical region)}
\ea\ee
The allowed range of $E$ is extremized at $|\mathbf{p}_\perp|=0$, giving
\begin{align}
    m-\frac{1}{2}\Delta V_i\leq E\leq -m+\frac{1}{2}\Delta V_i,
\end{align}
which exists only if $\Delta V_i>2m$. Additionally, for a fixed $E$ the allowed transverse momenta are bounded by $|\mathbf{p}_\perp|\leq \sqrt{(\Delta V_i/2-|E|)^2-m^2}$. 

There are two standard methods to calculate the rate of non-perturbative particle production in an external background. One method is to link the vacuum-persistence probability in the presence of an external source to the imaginary part of the effective action which is obtained by computing loop diagrams \cite{Fedotov:2022ely,Gelis:2015kya,Fedotov:2022ely,Koers:2004pj,nikishov1970pair,Maroto:1998zc}. The second approach, which we call the Bogoliubov method, is to link the pair-production probability in a spacetime patch to scattering amplitudes of a particle on a static potential barrier \cite{Fedotov:2022ely,Gelis:2015kya,Gavrilov:2015yha,Nikishov:2002ez,Chervyakov:2009bq,Chervyakov:2011nr,1999PhR...315...41D,1999PhR...315...41D,Gavrilov:2015zem,Kim:2009pg}. We will follow the Bogoliubov method and, for completeness, we briefly outline in the~Section~\ref{sec:pairproductionderivation} how $N_i(E,\bm{p}_\perp)$ arises in the Bogoliubov method. Full derivations can be found in standard treatments of the Dirac equation in pure-vector external backgrounds. For those who are not interested in the details, we provide here a quick sketch of the derivation
\begin{enumerate}
    \item \textbf{Asymptotic modes.} After reducing the Dirac equations to the scalar Schrodinger-like form, Eq.~\eqref{eq:reducedDirac}, the Sauter potential yields an equation exactly solvable in terms of hypergeometric functions. Because the Sauter potential asymptotes to constant values far from the step, $V_i(z\rightarrow \pm\infty)\rightarrow \pm |c_i|\Delta V/2$, one can define in and out plane-wave solutions with well-defined longitudinal momenta $p_{i,z}^\pm$. 
    \item \textbf{Canonical quantization and Bogoliubov coefficients.} Matching the exact solution to the asymptotic plane waves gives reflection and transmission amplitudes. These amplitudes determine the Bogoliubov coefficients relating the creation and annihilation operators in the in and out basis.
    \item \textbf{Mean number of produced pairs.} For a given $(E,\bm{p}_\perp)$, the expectation value of the produced pair number $N_i$ is related to the Bogoliubov coefficients, yielding  Eq.~\eqref{eq:Ni}.
\end{enumerate}

\section{Bound Neutrino Depletion}
\label{s:depletion}

\subsection{Absorption}
Since gradient produced neutrinos are extremely low-energy, typical weak cross-sections which scale as $\sigma_{\rm weak}\sim G_F^2E_\nu^2\sim 10^{-54}\text{ cm}^2$, are highly suppressed. There are, however, processes whose cross sections remain significant at low energies. A NS can build up free energy in several forms, including departures from chemical equilibrium and phase-conversion free energy. Gradient-produced neutrinos can then provide the necessary quantum numbers to catalyze exothermic reactions that release the free energy. For such reactions, the cross section can be set by the energy released $Q$ rather than the small incoming neutrino energy $E_\nu$.

For concreteness, consider as an example the process $\nu_es\rightarrow ue^-$ whose free-space cross-section is $\sigma_{\nu_e s}\sim G_F^2\theta_{\rm C}^2m_s^2\sim 2\times 10^{-41}\text{ cm}^2$, for Cabibbo angle $\theta_{\rm C}=0.23$ and strange quark mass $m_s=100\MeV$. Assuming the strange-quark number density is $n_s\sim 1\text{ fm}^{-3}$, the neutrino-absorption mean free path is naively $(n_s\sigma_{\nu_e s})^{-1}\sim 0.5\text{ m}$. Accounting for Pauli blocking of the final-state $u$ and $e^-$ increases the mean free path by a factor of $(\mu_{u}/T_{\rm core})(\mu_e/T_{\rm core})\sim 10^8$, for $\mu_u\sim \mu_e\sim 100\MeV$ and $T_{\rm core}\sim 10\keV$, yielding $\lambda_{\nu_e n}\sim 5\times 10^4\text{ km}$ \cite{1982AnPhy.141....1I}. If the NS is in a near chemical equilibrium, each absorption turns the neutrino's energy $E_\nu\sim \Delta V$, which originates from matter gradient, into kinetic energy carried by $u$ and $e^-$ which is rapidly converted into heat through electromagnetic processes. Thus, neutrino absorption effectively converts mechanical energy into thermal energy.

In an old and cold NS, chemical-equilibrating weak processes are extremely slow. For instance, the volume-integrated rate of chemical equilibration through strange-quark Urca processes, $s\rightarrow ue^-\bar{\nu}_e$ and $ue^-\rightarrow s\nu_e$, is approximately $\Gamma_{s,\rm Urca}\sim 10^{40}\text{ s}^{-1}(T_{\rm core}/\keV)^5(\delta\mu/T_{\rm core})$, which yields a relaxation time of $n_nR^3/\Gamma_{s,\rm Urca}\sim 10^{23}\text{ yr}(T_{\rm core}/\keV)^{-5}(\delta\mu/T_{\rm core})^{-1}$. Hence, a significant chemical-potential imbalance, {\it e.g.}, $\delta\mu=\mu_s-\mu_u-\mu_e-Q_{\rm vac}\neq 0$ with $Q_{\rm vac}=m_s-m_u-m_e$, is generically expected. Such an imbalance can both increase the energy deposited to NS matter per absorption from $E_\nu$ to $\delta \mu$ and increase the final-state phase space.  For $\delta\mu\gg \Delta V,T_{\rm core}$, the Pauli-blocking factor for the mean free path becomes less severe, $(\mu_u/\delta\mu)(\mu_e/\delta\mu)$. Such $\delta \mu$ can arise rotochemically: as a spinning-down NS contracts, the \textit{equilibrium} densities of $s,u,e^-$ shift by different amounts, while the \textit{actual} $s,u,e^-$ densities lag behind their equilibrium values, generating $\delta \mu\neq 0$ \cite{Reisenegger:1994be}. Rotochemical imbalances at the level of $\delta\mu\lesssim1\MeV$ is reasonable depending on the NS state \cite{2010A&A...522A..16G,Fernandez:2005cg} and can reduce the neutrino absorption mean free path by many orders of magnitude relative to the $\delta\mu=0$ case. The heating catalyzed by gradient-produced neutrinos may dominate over standard rotochemical heating from Urca processes \cite{1996ApJ...468..819C} for some $T_{\rm core}$ and $\delta\mu$ ranges. 

If neutrino absorption occurs, the released particles can heat up the NS interior. The heating rate can be estimated as
\begin{align}
    \dot{H}_{\nu,\rm abs}\sim  \frac{\dot{N}_\nu^{\rm grad}}{1+\lambda_{\nu,\rm abs}\dot{N}_\nu^{\rm grad}/N_\nu^{\rm deg}}\times(\Delta V+\delta \mu).
\end{align}
For the benchmark strange quark phase,
\begin{align}
    \lambda_{\nu,\rm abs}\sim 0.5\text{ m}\times \text{max}\left(1,\frac{100\text{ MeV}}{T_{\rm core}+\Delta V+\delta\mu}\right)^{2}.
\end{align}
Depending on the state of the NS interior, other similar exothermic neutrino absorption channels may exist, {\it e.g.}, $\nu_e d\rightarrow ue^-$ in quark matter and $\nu_e\Lambda\rightarrow pe^-$ and $\nu_e\Sigma^-\rightarrow ne^-$ in hyperonic matter \cite{1985ApJ...293..470G,Glendenning:1991es,Clevinger:2022xzl}. In an NS matter with a large pairing gap ({\it e.g.}~MeV), gradient-produced neutrino absorption can become important when there is sufficient build-up of chemical-potential imbalance  \cite{Reisenegger:1996ir,2010A&A...521A..77P,Yanagi:2019vrr}.

\subsection{Coherent Scattering}

Consider as an example baseline neutrino scattering on relativistic quarks in an unpaired deconfined matter. A crude estimate for the transport mean free path accounting for Pauli blocking yields a very large value, $\lambda_{\nu q}\sim [G_F^2E_\nu^2 n_q (T_{\rm core}/\mu_q)]^{-1}\sim 2\times 10^{13}\text{ km}(T_{\rm core}/10\keV)^{-1}$ for $E_\nu\sim 10\eV$, $n_q\sim 1\text{ fm}^{-3}$ and $\mu_q\sim 100\MeV$. However, gradient-produced neutrinos have de Broglie wavelengths that are typically much longer than the interparticle spacing of the NS medium. In this regime, the neutrinos do not probe individual constituents, but instead scatter coherently against collective excitations with a differential cross-section \cite{Bedaque:2018wns,Schuetrumpf:2019hqe}
\begin{align}
    \frac{d\sigma}{d\Omega}=\left(\frac{d\sigma}{d\Omega}\right)_{\rm free}\times S_V(q),
\end{align}
where $q$ is the spatial-momentum transfer and $S_V(q)$ is the (vector) static structure function of the medium, which can in principle enhance the neutrino cross section significantly relative to the free-space value.

To illustrate the degree to which long-range correlations can enhance neutrino scattering, suppose that the density fluctuations of NS matter have correlation lengths of order $\xi$ and are uncorrelated at larger length scales. The scattering amplitudes add up within a coherent volume, yielding  $S(q\rightarrow 0)\sim n_T\xi^3$. The maximal coherent enhancement is achieved when $\xi\sim E_\nu^{-1}\sim 10^{7}\text{ fm}$. To have an observable cooling effect, it suffices to have a mean free path $\lesssim 10^5\text{ km}$, which can be achieved with $\xi\gtrsim 100\text{ fm}$. Fluctuations may also appear in the form of large but rare (low filling fraction) blobs, in which case  of large coherent enhancement of each blob can compensate for the blobs' rareness. Whether long-range correlations occur in NS interiors is highly model dependent. Most neutrino-opacity studies focus on $\mathcal{O}(10)\MeV$ thermal neutrinos which probe the length scale of 10 fm, leaving NS fluctuations at larger length scales largerly unexplored. Potential sources include collective excitations \cite{Cermeno:2016olb} such as phonons, plasmons, and superfluid Goldstone modes, and various fluid phenomena such as convection and turbulence \cite{de2006hydrodynamic,schmitz1988fluctuations,kirkpatrick1982light,nicolis1971fluctuations,doyon2023emergence,1997Natur.390..262V}. In addition, we note that many proposals for detecting the C$\nu$B rely on coherent enhancements. Similar processes may enhance neutrino scattering or absorption in NS environments; see~Supplemental~Material~\ref{sec:dd}.

\subsection{Beyond the Standard Model}

BSM neutrino physics can change the fate and observability of gradient-produced neutrinos by opening new escape channels and/or enhancing their interactions in NS matter. New effective operators \cite{Babu:2019iml,Wise:2018rnb} or BSM electromagnetic properties \cite{Giunti:2014ixa,Broggini:2012df,Bell:2005kz,Giunti:2008ve,Akhmedov:2018wlf} can amplify neutrino interaction rates at 10 eV energies, drastically modifying the neutrino mean free paths. The latter could in principle open a new pair-production channel in strong NS magnetic fields \cite{Adorno:2021xvj,Lee:2005aj,Gavrilov:2012aw}. Active-sterile neutrino mixing can provide another escape channel for gradient-produced neutrinos \cite{Kusenko:2009up,Acero:2022wqg}. Once an active neutrino converts to a sterile state, it no longer feels the confining matter potential and free-streams out. Whether such conversion is efficient depends on the sterile mass and mixing parameters, as well as MSW effects. If neutrinos have enhanced self-interactions, {\it e.g.}~via a new light mediator \cite{Babu:2019iml,Graham:2025gtd,Banerjee:2025nvs}, then $\nu\nu\rightarrow \nu\nu$ can rapidly redistribute the energies of the trapped population. Even if gradient production initially populates energies below the confining barrier, fast thermalization can create a Fermi-Dirac-like distribution with an $\mathcal{O}(1)$ high energy tail extending above the potential well. Because the escaping neutrinos in these sterile-neutrino and self-interaction cases are ultimately powered by the free energy stored in the background potential gradient, they do not directly drain the NS thermal energy. However, sustained escape may alter the density/composition profile of the NS. The re-equilibration of the NS matter toward its preferred stratification may draw on internal energy (thermal or chemical) and can indirectly affect the cooling history. Quantifying this backreaction requres a self-consistent treatments of neutrino transport and NS stratification evolution. Any BSM scenario discussed here must satisfy laboratory, astrophysical, and cosmological constraints from, {\it e.g.}, neutrino-oscillation experiments, supernova, stellar cooling, the CMB, and BBN \cite{Dent:2016wcr}. Importantly, most existing bounds are derived at energies $\gtrsim \keV-\MeV$ or at early universe temperatures $\gtrsim \MeV$, whereas gradient-produced neutrinos are born with $\sim 10\eV$ energies. Translating existing constraints to this low-energy regime is often nontrivial and highly model dependent.

\section{Directly Detecting Gradient-Produced Neutrinos}

\label{sec:dd}

Each NS can spontaneously produce neutrino-antineutrino pairs via the gradient-production mechanism at an effective rate $\dot{N}_\nu^{\rm grad}F\lesssim 2\times 10^{38}\text{ s}^{-1}$ depending on its core temperature, the strength and sharpness of the matter potential jump, and the degree of Pauli blocking and reprocessing in the NS. If absorption dominates over scattering, the escaping component can be neutrino-antineutrino asymmetric, which could serve as distinguishing feature. These gradient-produced (anti)neutrinos would contribute to the local (anti)neutrino flux at Earth. Below we give crude estimates for the local flux, highlighting three contributions: 
\begin{enumerate}
    \item \textit{Nearest NS}. The nearest known NS is $\sim 100$ pc away, but population estimates suggest that an as-yet undetected NS could plausibly lie as close as $10\text{ pc}$ away, yielding a contribution to the local (anti)neutrino flux of $\lesssim  0.02\text{ cm}^{-2}\text{s}^{-1}$.
    \item \textit{Aggregate Galactic contribution}. The Milky Way is expected to host $\sim 10^9$ NSs.  Assuming, conservatively, that most of them are $\sim 10\text{ kpc}$ away yields a contribution to the local (anti)neutrino flux of $\lesssim 20 \text{ cm}^{-2}\text{ s}^{-1}$. This is likely the dominant contribution to the local flux in our simple scaling estimates, though a more realistic calculation should weight by the NS spatial, age, and temperature distributions as in, \textit{e.g.}, Ref.~\cite{Nguyen:2023czp}.
    \item \textit{Extragalactic NSs}. There are about $5\times 10^{-5}\rho_{\rm crit}H_0^{-3}/M_\odot\sim 5\times 10^{17}$ neutron stars in the observable universe \cite{Fukugita:2004ee}. This amounts to a total (anti)neutrino production rate of $\lesssim  1\times 10^{56}\text{ s}^{-1}$. Over a Hubble time $H_0^{-1}$, this yields a diffuse background with a present-day flux of $\phi\lesssim  0.1\text{ cm}^{-2}\text{s}^{-1}$, \textit{i.e.}, typically below the Galactic contribution in our benchmark assumptions.
\end{enumerate}
Gradient-produced neutrinos emerge from an NS with typical energies of $E_{\nu,\rm dec}$ ranging from $\Delta V\sim 10-100\eV$ up to $3T_{\rm core}\sim \keV-\MeV$ (for small scattering mean free paths and in hotter phases). Importantly, neutrinos from early/hot epochs may exist even when their impact on the cooling curve is subdominant to standard thermal neutrino emission and photon cooling. During those epochs, the gradient-produced neutrinos can have much smaller scattering/absorption mean free paths, and consequently, their production is less Pauli blocked. Relatedly, we note that gradient-produced neutrinos can be orders of magnitude less energetic than thermally produced ones in early stages, and so the former can exceed the latter in terms of \textit{number} flux even if the former has a smaller \textit{energy} flux. In the most optimistic case, the Milky Way contribution is comparable to the $\sim 10\,{\rm eV}$ neutrino background reported in \cite{Vitagliano:2019yzm}, which is expected to be dominated by the Sun's thermal emission. In principle, these backgrounds could be disentangled based on directions, spectral shape, and neutrino-antineutrino asymmetry.

The aforementioned $\gtrsim 10\eV$ (anti)neutrino backgrounds can in principle be probed with a PTOLEMY-like experiment \cite{Chacko:2018uke,McKeen:2018xyz,Reig:2019sok,Long:2014zva,Cocco:2007za,Betts:2013uya,PTOLEMY:2018jst} or other low-threshold detectors such as Borexino \cite{Bauer:2022lri,Borexino:2008gab}. PTOLEMY relies on neutrino capture on $\beta$-decaying nuclei $\nu_e+(A,Z)\rightarrow e^-+(A,Z+1)$ and aims at detecting the primordial C$\nu$B with a tritium target of mass $M_T=100\text{ g}$. The same setup can be used to detect electron neutrinos with energy much higher than $0.1\meV$. The cross section for electron capture on tritium is $\sigma\approx 4\times 10^{-45}\text{ cm}^2$ for $E_\nu\ll \keV$. This yields an event rate of $\sim 9\text{ yr}^{-1}$ for the C$\nu$B and $2\times 10^{-9}\text{ yr}^{-1}$ for Galactic gradient-produced neutrinos. That said, the optimal detector for gradient-produced neutrinos need not match PTOLEMY's priorities. One could imagine, for instance, trading energy resolution for a vastly larger target mass, higher acceptance, or alternative isotopes to increase the total rate. Furthermore, solid-state detectors are pushing toward sensitivity to eV-scale energy depositions \cite{Baxter:2025odk}. Low-energy neutrinos could also be probed indirectly via cosmic-ray upscattering into the energy range accessible to IceCube-like detectors \cite{Herrera:2026pzj} and via the resulting electromagnetic emission \cite{Herrera:2026bie}. Escaping Gradient-produced neutrinos at higher energies would benefit from larger interaction cross sections, compensating their smaller flux relative to the C$\nu$B. More broadly, proposals exploiting coherent enhancements or collective effects for light-particle detection may be repurposed for targeting gradient-produced neutrinos \cite{Akhmedov:2018wlf,Arvanitaki:2024taq,Galanis:2025amc,Bauer:2022lri,Berlin:2024ewa,Chang:2022gcs,Liu:2025jlx}.

\section{Thermal-neutrino luminosities}

For the benchmark parameters in Eq.~\eqref{eq:benchmarks}, we take \cite{Yakovlev:2004iq,Page:2004fy}
\begin{align}
    L_{\nu\bar{\nu}}^{\rm th}&= a\left(\frac{T_{\rm core}}{100\eV}\right)^b, \label{eq:Lth}
\end{align}
where $a$ and $b$ are model-dependent quantities, and $T_{\rm core}$ and $T_{\rm sur}$ are related via Eq.~\eqref{eq:envelope}. Fig.~\ref{fig:LvsTsur} shows that $L_{\nu\bar{\nu}}^{\rm grad}$ can exceed both $L_{\gamma}$ and $L_{\nu\bar{\nu}}^{\rm th}$ for the surface-temperature range $T_{\rm sur}\sim 10^3-10^5\text{ K}$, which spans from optical to UV. Although we are mainly interested in NS temperatures where thermal neutrino emission is negligible, we show for illustration purposes three representative thermal-neutrino luminosities relevant in different NS models: (1) $n\rightarrow pe\bar{\nu},\,pe\rightarrow n\nu$ (Urca) with $a\sim 1\times 10^{27}-3\times 10^{28}\text{ erg/s}$ and $b=6$, (2) $nN\rightarrow pNe\bar{\nu},\, pNe\rightarrow nN\nu$ (MUrca) with $N\in (n,p)$, $a\sim 8\times 10^{15}-2\times 10^{17}\text{ erg/s}$ and $b=8$, and (3) processes that occur in quark matter $d\rightarrow ue\bar{\nu},\, ue\rightarrow d\nu$ (quark) with $a\sim 10^{24}-10^{25}\text{erg/s}$ and $b=6$ \cite{Yakovlev:2004iq}. For simplicity, we take the thermal emission to occur in the entire volume of the NS of radius $R$.

\section{Pair-Production Rate Calculation (Bogoliubov Method)}\label{sec:pairproductionderivation}
With the Sauter potential $V(z)=(\Delta V/2)\tanh(z/l)$, the Dirac equation is exactly solvable in terms of hypergeometric functions, as done in, {\it e.g.}, Refs.~\cite{Chervyakov:2009bq,Chervyakov:2011nr,1999PhR...315...41D}. The set of independent solutions admit free plane-wave limits at $z\rightarrow\pm \infty$, which define natural in/out particle and antiparticle states. Below we describe how these asymptotic states are related and how their relations determine the mean pair-production number $N_i(E,\bm{p}_\perp)$ quoted in the previous section. Throughout this section, we take $c_{L,R}>0$ for simplicity and without loss of generality, as explained in the previous section.

\subsection{Asymptotic modes}
To distinguish the multiple independent solutions of the Dirac equation defined by their asymptotic plane-wave limits, we adopt the notation $^{r}\psi_{i,s}$ and $_{r}\psi_{i,s}$, where $i=L,R$ refers to the chiral component and $s=\pm 1$ is the spin label (eigenvalues of $\sigma_3$), as before. We also introduce the index $r = \pm$, whose value refers to $\pm\hat{z}$ travel \textit{direction} of the wave and whose location (subscript $_r$ vs superscript $^r$) labels $z\rightarrow \pm \infty$ asymptotic spatial \textit{location} in which the plane-wave solution applies. Specifically, solutions with an upper index $^{r}$ travel in the $r\hat{z}$ direction and asymptotically become plane waves at $z=+\infty$:
\be\ba \label{eq:}
^{\pm}\psi_{i,s} ( z\to + \infty) \propto e^{\pm i p_{i,z}^{+} z} .
\ea\ee
while solution with a lower index $_{r}$ travel in the $r\hat{z}$ direction and asymptotically becomes plane waves at $z=-\infty$:
\be\ba \label{eq:}
_{\pm}\psi_{i,s} ( z\to - \infty) \propto e^{\pm i p_{i,z}^{-} z}.
\ea\ee
For given $(E,\bm{p}_\perp,s)$, each pair $(^+\psi,^-\psi)$ and $(_+\psi,_-\psi)$ forms a complete basis of solutions. Because the equation is linear in $\psi$, these two bases are related by a linear transformation. In the standard scattering-problem interpretation with incoming wave from $z=-\infty$ moving toward $+\hat{z}$, reflected back to $z=-\infty$, and transmitted to $z=+\infty$, one may write
\be\ba \label{eq:IRleft}
\underbrace{^+\psi_{i,s} (z)}_{\rm transmitted}
=
\underbrace{I_i \,  {_+}\psi_{i,s}(z)}_{\rm incoming}
 + \underbrace{R_i \, {_-}\psi_{i,s} (z)}_{\rm reflected},
\ea\ee
and, similarly, with incoming wave from $z=+\infty$ moving toward $-\hat{z}$, reflected back to $z=+\infty$, and transmitted to $z=-\infty$, 
\be\ba \label{eq:IRright}
\underbrace{^-\psi_{i,s} (z)}_{\rm transmitted}
=
\underbrace{-R_i^* {_+}\psi_{i,s} (z) }_{\rm reflected}
\underbrace{ - I_i^*  {_-}\psi_{i,s}(z)}_{\rm incoming} .
\ea\ee
Here $I_i$ and $R_i$ are the transmission and reflection amplitudes for the corresponding single-particle scattering problem. For the Sauter potential these amplitudes are known analytically in terms of Gamma functions
\be\ba \label{eq:IR}
I_i = \frac{{^+}N_{i,s}}{{_+}N_{i,s}} g \left(-\frac{1}{2} s  l \Delta V_i ,\,\frac{1}{2} l p_{i,z}^{-} ,\,\frac{1}{2}l p_{i,z}^{+} \right) ,\quad 
R_i =
 \frac{{^+}N_{i,s}}{{_-}N_{i,s}} h \left(-\frac{1}{2}s l \Delta V_i ,\,\frac{1}{2}l p_{i,z}^{-} ,\,\frac{1}{2}l p_{i,z}^{+} \right) ,
\ea\ee
where the functions $g$ and $h$ are defined as
\be\ba \label{eq:trans-coef}
& g(\lambda,\mu,\nu)=\frac{\Gamma(1-2 i \nu) \Gamma(-2 i \mu)}{\Gamma[i(\lambda-\nu-\mu)] \Gamma[1-i(\lambda+\nu+\mu)]}, \\
& h(\lambda,\mu,\nu)=\frac{\Gamma(1-2 i \nu) \Gamma(2 i \mu)}{\Gamma[i(\lambda-\nu+\mu)] \Gamma[1-i(\lambda+\nu-\mu)]}.
\ea\ee
The normalization factors $^rN_{i,s}$ and $_rN_{i,s}$ are fixed by unit flux normalization of the Dirac current in the asymptotic regions
\be\ba \label{eq:normalizations}
{^r}N_{i,s} = \left| 2 p_{i,z}^{+} \left( p_{i,0}^{+} - s\, r\, p_{i,z}^{+}\right)\right|^{-\frac{1}{2}},\quad 
{_r}N_{i,s} = \left|  2 p_{i,z}^{-} \left( p_{i,0}^{-} - s\, r\, p_{i,z}^{-}\right) \right|^{-\frac{1}{2}}.
\ea\ee
For this choice of normalization, the reflection and transmission probabilities are 
\be\ba \label{eq:rt}
r_i = \frac{|R_i|^2}{|I_i|^2}, \quad t_i =  \frac{1}{|I_i|^2}.
\ea\ee
In the supercritical/Klein regime, where the potential step $\Delta V$ is strong enough that positive-frequency states on one side connect to negative-frequency states on the other, current conservation implies $r_i-t_i=1$, and so $r_i>1$. This seemingly unphysical reflection coefficient is the source of confusion that leads to the Klein paradox, whose resolution requires a many-body interpretation involving pair production \cite{1999PhR...315...41D}, which we discuss next.

\subsection{Canonical quantization and Bogoliubov transformation}

The connection to particle creation is made by canonically quantizing the field operators $\Psi_{L,R}$ and expanding them in a complete set of mode functions. One may choose either the in basis (defined by incoming waves at time $t\rightarrow -\infty$) and the out basis (defined by outgoing waves at time $t\rightarrow+\infty$). In the supercritical/Klein regime, only particles have real longitudinal momenta on one side of the potential and only antiparticles have real longitudinal momenta on the other. Correspondingly, the mode functions with subscript $_r$(superscript $^r$), namely $_\pm\psi_i$($^\pm\psi_i$), which asymptotically become free plane waves at $z\rightarrow -\infty$ ($z\rightarrow+\infty$) can be interpreted as particle(antiparticle) solutions. The two different ways (in/out basis) of expanding $\Psi_{L,R}$ lead to the following relation
\be\ba \label{eq:}
\Psi_i (x) = \sum_{E,\bm{p}_{\perp},s} \left[ a^{{\rm in}}_{i,s} \, {_{+}}\psi_{i,s} (x) + (b_{i,s}^{\rm in})^{\dagger}  \, {^{+}}\psi_{i,s} (x)  \right] =\sum_{E,\bm{p}_{\perp},s} \left[ a^{{\rm out}}_{i,s} \, {_{-}}\psi_{i,s}  (x) + (b_{i,s}^{\rm out})^{\dagger} \, {^{-}}\psi_{i,s}  (x)  \right],
\ea\ee
where the operator $a$($b$) annihilates particles(antiparticles) in the specified time limits (in/out) and obeys the standard anticommutation relations, the $(E,\bm{p}_\perp)$ dependence is suppressed for readability, the phase-space integration is shortened as $\sum_{E,\bm{p}_{\perp},s} \equiv\sum_s A T \int \frac{d E d^2 \bm{p}_{\perp}}{(2\pi)^3}$, $A = \int dx dy$ is a transverse area, and $T$ is a time interval. Using the linear relations between mode functions, Eqs.~\eqref{eq:IRleft} and \eqref{eq:IRright}, the above equality leads to the Bogoliubov transformation between the in and out operators
\be\ba \label{eq:Bogoliubov}
& a^{{\rm in}}_{i,s} = -\frac{I}{R} a^{{\rm out}}_{i,s} -\frac{1}{R} (b_{i,s}^{\rm out})^{\dagger}  ,\\
& (b_{i,s}^{\rm in})^{\dagger}  = \frac{1}{R} a^{{\rm out}}_{i,s} - \frac{I^*}{R} (b_{i,s}^{\rm out})^{\dagger} .
\ea\ee
As we will see next, this mixing of $a$ and $b^\dagger$ entails particle production.

\subsection{Mean number of produced pairs}
We define the in and out vacua by
\begin{align}
    a^{\rm in}|0_{\rm in}\rangle=b^{\rm in}|0_{\rm in}\rangle=0,\quad a^{\rm out}|0_{\rm out}\rangle=b^{\rm in}|0_{\rm out}\rangle=0,
\end{align}
where $(E,\bm{p}_\perp,s,x)$ dependencies are suppressed. For a given phase-space patch at $(E,\bm{p}_\perp,x)$, the expectation value of the number of produced pairs is given by
\be\ba \label{eq:Nicalculation}
 {N}_i ( {E,\bm{p}_{\perp}} )= |\langle 0_{\rm out} | b^{{\rm out}}_{i,s} a^{{\rm out}}_{i,s} | 0_{\rm in}\rangle |^2 
 =  t_i ( {E,\bm{p}_{\perp}} )\, P_i ( {E,\bm{p}_{\perp}} ) = \frac{t_i(E,\bm{p}_{\perp})}{r_i (E,\bm{p}_{\perp})}.
\ea\ee
The second equality follows from the Bogoliubov transformation Eq.~\eqref{eq:Bogoliubov} and ${P}_i ( {E,\bm{p}_{\perp}} )\equiv |\langle 0_{\rm out} | 0_{\rm in} \rangle|^2$ is the vacuum persistence probability. The last equality follows from the usual fermionic anticommutation relations which imply $P_i+N_i=1$ (i.e. the Pauli exclusion principle) and current conservation $r_i(E,\bm{p}_{\perp}) = 1+t_i(E,\bm{p}_{\perp})$. The pair-production rate per unit transverse area in sector $i$ is then obtained by integrating over phase-space patches
 \be\ba
 \dot{n}_{i,\perp} \equiv  \frac{N_{i} /T}{A} = \int \frac{d E d^2 \bm{p}_{\perp}}{(2\pi)^3} N_i (E,\bm{p}_{\perp}).
\ea\ee
Finally, evaluating the $t_i/r_i$ in Eq.~\eqref{eq:Nicalculation} using Eqs.~\eqref{eq:IR}, \eqref{eq:trans-coef},\eqref{eq:normalizations}, \&\eqref{eq:rt} yields
\begin{align}\label{eq:appgenrateperarea}
 N_i (E,\bm{p}_{\perp})
  =  \frac{\sinh \left( \pi l p_{i,z}^{+} \right) \sinh \left(  \pi l  p_{i,z}^{-} \right)}{ \sinh \frac{\pi l}{2} \left(\Delta V_{i} +  p_{i,z}^{+} -  p_{i,z}^{-}  \right) \sinh  \frac{\pi l}{2} \left( \Delta V_i - p_{i,z}^{+} +  p_{i,z}^{-}  \right)},
\end{align} 
which reproduces Eq.~\eqref{eq:Ni}.

\section{Gradual, sharp limits, and the spectra}

\label{sec:limits}
In this section, we show how to extract useful limiting forms of the Sauter-potential pair-production rate per unit transverse area, Eqs.~\eqref{eq:ndotperp} and $\eqref{eq:Ni}$, in the limits $l\gg \Delta V_i^{-1}$ (gradual) and $l\ll \Delta V_i^{-1}$ (sharp). We also discuss the energy spectra of the produced excitations.

\subsection{Gradual limit ($l\gg \Delta V_i^{-1}$)}
The pair-production rate per unit transverse are, Eq.~\eqref{eq:ndotperp}, can be rewritten as
\be\ba \label{eq:}
\dot{n}_{\perp} 
&= \frac{1}{ (2 \pi)^2} \sum_{i=L,R}
 \int_0^{\Delta V_i/2-m} d E \int_0^{(\Delta V_i/2-E)^2-m^2} d p_{\perp}^2 
\, 
\frac{\cosh \pi l \left(p_{i,z}^- + p_{i,z}^+\right) - \cosh \pi l  \left(p_{i,z}^- - p_{i,z}^+\right)}
{ \cosh  \pi l \Delta V_i - \cosh \pi l  \left(p_{i,z}^- - p_{i,z}^+\right)},
\ea\ee
where the phase space integration limit has been derived in~Eq.~\eqref{eq:phasespace}. Because in the gradual limit $l\Delta V\gg 1$, the $\cosh ( \pi l \Delta V_i)$ in the denominator is exponentially large, and so the integrand is non-vanishing only when an exponential in the numerator compensates that of $\cosh ( \pi l \Delta V_i)$  in the denominator. This occurs when  $p_{i,z}^- + p_{i,z}^+ \approx \Delta V_i$ ({\it i.e.}~$p_{i,z}^- - p_{i,z}^+ \approx 2 E$) for the first term in the numerator. The second term in the numerator requires $p_{i,z}^- - p_{i,z}^+ \approx \Delta V_i$, however this lies outside of the supercritical/Klein regime. Expanding $p_{i,z}^- + p_{i,z}^+$ around $\Delta V_i$ for small transverse mass $m_\perp^2\equiv p_\perp^2 + m^2 \ll (\Delta V_i/2+E)^2,(\Delta V_i/2-E)^2$ gives
\be\ba \label{eq:}
p_{i,z}^- + p_{i,z}^+ \approx \Delta V_i - \frac{m_\perp^2\Delta V_i/2}{ (\Delta V_i/2)^2 - E^2 }.
\ea\ee
Hence, for $l \gg \Delta V_i^{-1}$, the integral simplifies as
\be\ba \label{eq:}
\dot{n}_{\perp} 
&\approx 
 \frac{1}{ (2 \pi)^2} \sum_{i=L,R}
 \int_0^{\Delta V_i/2-m} d E \int_{m^2}^{(\Delta V_i/2-E)^2} d m_{\perp}^2 
\, 
\exp \left[ 
- 
\frac{\pi m_{\perp}^2}{|c_i| g}\right] 
\approx \frac{1}{ 4 \pi^3 } l \sum_{i=L,R} | c_i\,g |^2 
e^{- \pi \frac{m^2}{|c_i| g}}\quad \text{(gradual limit)},
\ea\ee
where we have defined $g\equiv \Delta V/(2l)$. 

Since pair-production occurs predominantly in the transition region where the Sauter potential varies appreciably, one cannot derive the exact rate per unit volume from the rate per unit transverse area. An approximate rate per unit volume $\dot{n}$ can be obtained by dividing the obtained $\dot{n}_\perp$ by the approximate spatial extent of the transition region $\sim 2l$
\be\ba \label{eq:}
\dot{n} &\sim \frac{\dot{n}_{\perp}}{2 l} \approx
 \frac{1}{ 8 \pi^3 } \sum_{i=L,R} | c_i\,g |^2 
e^{- \pi \frac{m^2}{|c_i| g}}.
\ea\ee
The gradual limit is closely related to the Schwinger limit, which corresponds to taking the strict $\Delta V \to \infty, L \to \infty$ limits while keeping the ratio $g=\Delta V /(2L)$ fixed. In this Schwinger limit, the Sauter potential reduces for any finite $z$ to $V(z)=(\Delta V/2)\tanh(z/l)\rightarrow gz$. For a Dirac fermion of mass $m$ and charge $e$ in a constant and uniform electric field $E_{\rm EM}$ one has $c_L = c_R =c_V=1$ and $g=|e|E_{\rm EM}$. Then the above becomes 
\be\ba \label{eq:}
\dot{n}\sim \frac{1}{ 4 \pi^3 } | e  E_{\rm EM} |^2 
e^{- \frac{\pi m^2}{|e|  E_{\rm EM}}}\quad\text{(Schwinger limit)},
 \ea\ee
which is the standard Schwinger result.

\subsection{Sharp limit ($l\ll \Delta V_i^{-1})$}

In the sharp limit, $l\ll \Delta V_i^{-1}$, the pair-production rate per unit transverse area, Eq.~\eqref{eq:ndotperp}, can be simplified as follows
\be\ba \label{eq:ndotperpsharp}
\dot{n}_{\perp} 
&=\sum_{i=L,R} \frac{\Delta V_i^3}{16 \pi^2} \int_0^{1-\epsilon } d x \int_{\epsilon_i^2}^{(1-x)^2} d y 
\,
\frac{\sinh \left( \pi l \Delta V_i \sqrt{(1 - x)^2- y }/2\right) \sinh \left( \pi L \Delta V_i \sqrt{(1 + x)^2 - y }/2 \right)}
{ \cosh ( \pi l \Delta V_i) - \cosh \frac{\pi l \Delta V_i}{2}  \left( \sqrt{(1+ x)^2- y } -  \sqrt{(1 - x)^2- y }  \right)} \\
&\approx  
 \sum_{i=L,R}\frac{\Delta V_i^3}{ 16 \pi^2} \int_0^{1-\epsilon_i } d x \int_{\epsilon_i^2}^{(1-x)^2} d y 
\, 
\frac{ \sqrt{ ( y - 1 - x^2 )^2 - 4 x^2} }
{ 2 + ( y - 1 - x^2 ) + \sqrt{ ( y - 1 - x^2 )^2 - 4 x^2}  } \\
& =  \sum_{i=L,R}
 \frac{\Delta V_i^3}{ (8 \pi)^2} \int_0^{1-\epsilon_i } d x 
\, 
\left[ 
 - 2 + (x^2 + 1) (3 - 2 x) + 2 \ln(1+x) - 2 x^2 (2 - x^2) \ln (1 + x^{-1})
\right] + \mathcal{O}(\epsilon^2) \\
&=  \sum_{i=L,R}
 \frac{16 \, \Delta V_i^3}{ 15 (8 \pi)^2}
 (  \ln 4 - 1 ) + \mathcal{O} (\epsilon^2),\quad\quad\quad\quad\quad\quad\text{(sharp limit)}
\ea\ee
where we have defined 
\begin{align}
    x\equiv \frac{2E}{\Delta V_i},\quad y\equiv \frac{4m_\perp^2}{\Delta V_i^2},\quad \epsilon_i \equiv \frac{2m}{\Delta V_i}.
\end{align}
In going to the second line, we took the small-argument ($l\Delta V_i\ll 1$) limits of the hyperbolic trigonometric functions. In going to the third line, we expanded in the parameter $\epsilon_i$ which is $\ll 1$ in the in the small-mass limit. Finally, we performed the remaining $x$ integral to arrive at the expression on the last line.

\subsection{Energy spectra of the produced excitations in the sharp limit}
\begin{figure}
    \centering
    \includegraphics[width=0.6\linewidth]{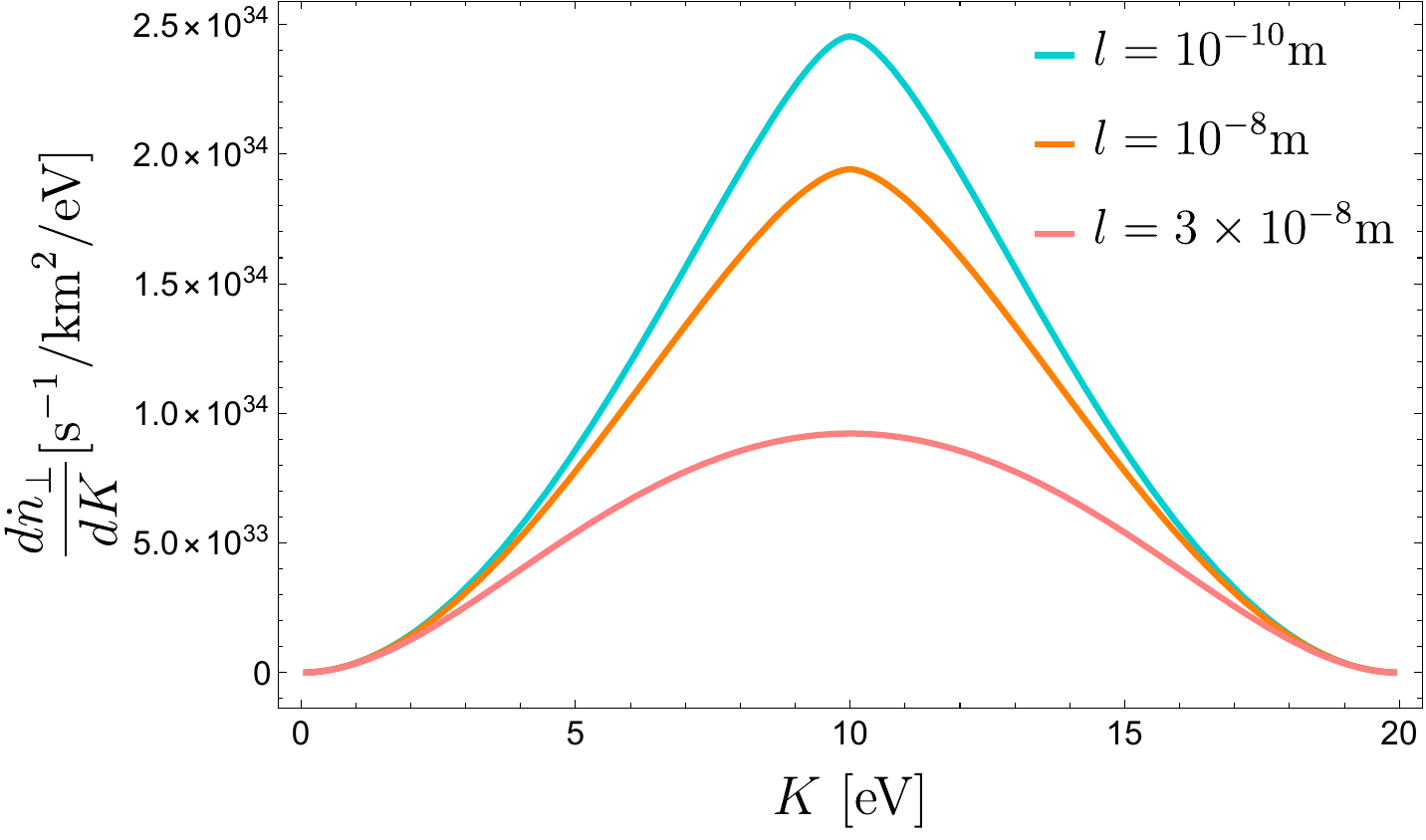}    
    \caption{Gradient-produced particle kinetic-energy spectra in the sharp limit ($l\ll \Delta V^{-1}$) for $\Delta V=20\eV$ and $c_V=c_A=1/2$. More precisely, the quantity being plotted on the vertical axis is the particle production rate per unit area per unit (asymptotic) kinetic energy $K=\sqrt{\bm{p}^2+m^2}-m$. }
    \label{fig:placeholder}
\end{figure}

The energy spectra of the produced particles and antiparticles can be found by performing the integral in the first line of Eq.~\eqref{eq:ndotperpsharp} over the transverse momentum $\bm{p}_{\perp}$ (\textit{i.e.~}$y$) only,
\be\ba \label{eq:spectrum}
\frac{d \dot{n}_{\perp}}{d E}
&=\sum_{i=L,R} \frac{\Delta V_i^2}{( 4 \pi)^2} \int_{\epsilon_i^2}^{(1-|x|)^2} d y 
\,
\frac{\sinh \left( \pi l \Delta V_i \sqrt{(1 - x)^2- y }/2\right) \sinh \left( \pi L \Delta V_i \sqrt{(1 + x)^2 - y }/2 \right)}
{ \cosh ( \pi l \Delta V_i) - \cosh \frac{\pi l \Delta V_i}{2}  \left( \sqrt{(1+ x)^2- y } -  \sqrt{(1 - x)^2- y }  \right)},
\ea\ee
where $x \equiv 2 E/\Delta V_i$, $-1 + \epsilon_i<x<1 - \epsilon_i$ in the supercritical/Klein regime, and the spectrum is even under $E \to -E$. In the sharp limit, the spectrum is approximately
\be\ba \label{eq:spectrumsharp}
\frac{d \dot{n}_{\perp}}{d E}
&\approx  \sum_{i=L,R}
 \frac{\Delta V_i^2}{ (8 \pi)^2} 
\left[ 
  (x^2 + 1) (3 - 2 |x|) -2 + 2 \ln(1+|x|) - 2 x^2 (2 - x^2) \ln (1 + |x|^{-1})
\right] + \mathcal{O} (\epsilon_i^2) \\
&\approx \sum_{i=L,R}
 \frac{\Delta V_i^2}{ (8 \pi)^2} 
 \begin{cases}
1 + 4 (2 E/\Delta V_i)^2 \log \left| 2 E /\Delta V_i \right|,\quad |E| \ll \Delta V_i/2 \\
    4 \,(\log4 -1 ) \big[ 1 - (2E/\Delta V_i)^2\big],\quad 
    E \to \Delta V_i/2 \\
 \end{cases}.
\ea\ee
As shown in the second line of~Eq.~\eqref{eq:spectrumsharp}, the spectrum peaks around $E=0$, and asymptotes to zero as $E \to \Delta V_i/2$. The above spectrum is only valid in the supercritical/Klein regime, which corresponds to $x \le 1 - \epsilon_i$ or $E \le \Delta V_i/2 - m$. In this regime, particles emerge as asymptotically free excitations on one side of the potential and antiparticles on the other. The particle propagates toward the side where a positive-frequency mode is kinematically allowed (in our convention, this is the $z<0$ side), while antiparticles propagate toward the opposite side ($z>0$). Because the potential is static, the frequency $E$ (at production) is a conserved quantity. The asymptotic kinetic energy is obtained by subtracting the appropriate potential energy in the asymptotic region $V_i\rightarrow \pm\Delta V_i/2$. For a produced particle (which becomes asymptotically free at $z\rightarrow-\infty$) the asymptotic kinetic energy is
\begin{align}
K_{i} &\equiv \sqrt{\bm{p}_i(z\rightarrow-\infty)^2 + m^2} - m = | E - V_i(z\rightarrow -\infty) | - m = E+\frac{1}{2} \Delta V_i - m.
\end{align}
Similarly, we find that for an antiparticle (which becomes asymptotically free at $z\rightarrow +\infty$) the asymptotic kinetic energy is given by $-E+\Delta V_i/2-m$. 
It is arguably more useful to present the spectrum in terms of $K$ rather than $E$. While $d\dot{n}_\perp/dK=d\dot{n}_\perp/dE$, the allowed range of $K$ is shifted relative to that of $E$.

In~Fig.~\textcolor{blue}{S}\ref{fig:lumi}, we show the \textit{particle} kinetic-energy spectrum for $\Delta V=20\eV$ and $c_V=c_A=1/2$ (i.e., the benchmark values assumed in the main text), and various values of the transition width $l$ satisfying $l\ll \Delta V^{-1}$. It can be read from the discussion around both Eq.~\eqref{eq:spectrumsharp} and Fig.~\textcolor{blue}{S}\ref{fig:lumi} that most of the fermions acquire a kinetic energy of order the potential jump strength, $K\sim \Delta V/2$, consistent with the classical picture in which the pair is created near the interface and then accelerated by the potential gradient before emerging into the asymptotic region. Furthermore, \textit{all} the produced particles which moves toward the $z<0$ side have kinetic energies $K$ less than the potential step strength $|\Delta V|$, as $E \le \Delta V/2 - m$; see the discussion around Eq.~\eqref{eq:phasespace}. For the NS application, this means that all the produced neutrinos have total energies that lie below the barrier height, and so they are kinematically forbidden from escaping the NS region enclosed by the potential-jump interface.